\documentclass[12pt,a4paper]{article}
\usepackage{graphicx}
\usepackage{subeqnarray}
\begin{document}
%
%
%
%
\newenvironment{lefteqnarray}{\arraycolsep=0pt\begin{eqnarray}}
{\end{eqnarray}\protect\aftergroup\ignorespaces}
\newenvironment{lefteqnarray*}{\arraycolsep=0pt\begin{eqnarray*}}
{\end{eqnarray*}\protect\aftergroup\ignorespaces}
\newenvironment{leftsubeqnarray}{\arraycolsep=0pt\begin{subeqnarray}}
{\end{subeqnarray}\protect\aftergroup\ignorespaces}
\newcommand{\diff}{{\rm\,d}}
\newcommand{\pprime}{{\prime\prime}}
\newcommand{\szeta}{\mskip 3mu /\mskip-10mu \zeta}
\newcommand{\srho}{\mskip 3mu /\mskip-10mu \rho}
\newcommand{\sr}{\mskip 3mu /\mskip-9mu r}
\newcommand{\sR}{\mskip 3mu /\mskip-11mu R}
\newcommand{\sV}{\mskip 3mu /\mskip-10mu V}
\newcommand{\sP}{\mskip 3mu /\mskip-10mu p}
\newcommand{\sPM}{\mskip 3mu /\mskip-11mu P}
\newcommand{\sT}{\mskip 3mu /\mskip-9mu T}
\newcommand{\FC}{\mskip 0mu {\rm F}\mskip-10mu{\rm C}}
\newcommand{\appleq}{\stackrel{<}{\sim}}
\newcommand{\appgeq}{\stackrel{>}{\sim}}
\newcommand{\quadr}{\overline\sqcup}
\newcommand{\Int}{\mathop{\rm Int}\nolimits}
\newcommand{\Nint}{\mathop{\rm Nint}\nolimits}
\newcommand{\arcsinh}{\mathop{\rm arcsinh}\nolimits}
\newcommand{\range}{{\rm -}}
\newcommand{\displayfrac}[2]{\frac{\displaystyle #1}{\displaystyle #2}}
%
%
\title{On the desert between neutron star and black hole remnants}

\author{
 {R.~Caimmi}\footnote{
{\it Physics and Astronomy Department, Padua University,
Vicolo Osservatorio 3/2, I-35122 Padova, Italy.   Affiliated up to September
30th 2014.  Current status: Studioso Senior.   Current position: in retirement
due to age limits.}\hspace{50mm}
email: roberto.caimmi@unipd.it~~~
fax: 39-049-8278212}
\phantom{agga}}

\maketitle
\begin{quotation}
\section*{}
\begin{Large}
\begin{center}

 Abstract

\end{center}
\end{Large}
\begin{small}

\noindent\noindent
The occurrence of a desert between neutron star (NS) and black hole (BH)
remnants is reviewed using a set of well-determined masses from different
sources.   The dependence of stellar remnants on the zero age main sequence
(ZAMS) progenitor mass for solar metallicity is taken from a recent
investigation and further effort is devoted to NS and BH remnants.   In
particular, a density parameter is defined and related to NS mass and radius.
Spinning BHs in Kerr metrics are considered as infinitely thin, homogeneous,
rigidly rotating disks in Newtonian mechanics.   Physical parameters for
nonrotating (TOV) and equatorial breakup (EQB) configurations are taken or
inferred from a recent investigation with regard to 4 NS and 3 quark star (QS)
physically motivated equation of state (EOS) kinds.   A comparison is
performed with counterparts related to nonrotating and maximally rotating BHs.
The results are also considered in the light of empirical
relations present in literature.   With regard to $J$-$M$ relation, EQB
configurations are placed on a sequence of similar slope in comparison to
maximally rotating BHs, but shifted downwards due to lower angular momentum by
a factor of about 3.6.   Under the assumption heavy baryons are NS
constituents and instantaneously undergo quark-level reactions, the energy
released (or adsorbed) is calculated using results from a recent
investigation.   Even if NSs exclusively host heavy baryons of the kind
considered, the total amount cannot exceed about 10\% of the binding energy,
which inhibits supernova explosions as in supramassive white dwarf (WD)
remnants or implosions into BH.   Alternative channels for submassive
($2\appleq M/m_\odot\appleq4$) BH formation are shortly discussed.   Whether
the desert between NS and BH remnants could be ascribed to biases and/or
selection effects, or related to lack of formation channels, still remains an
open question.   To this respect, increasingly refined theoretical and
observational techniques are needed.

\noindent

{\it keywords -
stars: evolution - stellar remnants - nucleosynthesis - neutron stars - black
holes.}
\end{small}
\end{quotation}

\section{Introduction} \label{intro}

Stellar remnants are ending configurations (if any) of related zero-age main
sequence (ZAMS) progenitors at the end of stellar evolution.   Leaving aside
brown dwarf (BD) stars, which might be considered as ending configurations by
themselves after a short period of D- and possibly Li-burning \cite{SBM11}, a
stellar remnant can be classified as white dwarf (WD), neutron star (NS)
including hybrid stars and quark stars (QS), black hole (BH).

WD progenitors are characterized by sufficiently low masses where nuclear
reactions are inhibited before Fe nucleosynthesis.   More specifically, ZAMS
progenitors within the mass range, $0.5\appleq M/m_\odot\appleq8$, end stellar
evolution as carbon-oxygen WDs.  Nonrotating WDs are unstable above the
Chandrasekhar mass limit, $M_{\rm Ch}\approx1.44\,m_\odot$, yielding a
supernova
explosion of type Ia e.g., \cite{BaS00}.   Ultramassive WDs could occur in
presence of fast rotation e.g., \cite{OBL66} \cite{OsB68} or strong
magnetic fields e.g., \cite{KuM12} \cite{DaM14}.

NS progenitors are characterized by sufficiently large masses where
nuclear reactions allow Fe nucleosynthesis.   More specifically, ZAMS
progenitors within the range, $8\appleq M/m_\odot\appleq120$, end stellar
evolution as either NSs after core collapse and supernova explosion of type
II, or BHs.   Nonrotating NSs are unstable above a threshold,
$M_{\rm TOV}\appgeq2m_\odot$, depending on nuclear matter equation of state
(EOS) e.g., \cite{LaP04} \cite{YaY13} \cite{OzF16} \cite{SaL17} \cite{YaY17}
\cite{RSW18} \cite{SLI18} \cite{SBH18}.   Ultramassive
NSs could occur in presence of fast rotation e.g., \cite{haa16} \cite{RMW18}
\cite{zha17} or strong magnetic fields e.g., \cite{lia16}.

BH progenitors lack a clear mass distinction with respect to NS progenitors.
According to a recent investigation \cite{RSO18}, the above mentioned ZAMS
progenitors co-exist within the mass range, $15\appleq M/m_\odot\appleq21$;
$25\appleq M/m_\odot\appleq28$; $60\appleq M/m_\odot\appleq120$; while only
the former takes place for $10\appleq M/m_\odot\appleq15$ and the only latter
for $21\appleq M/m_\odot\appleq25$; $28\appleq M/m_\odot\appleq60$.   The
occurrence of NS or BH remnants might be related to different extents of mass
loss during evolution, with the addition of neutrino transport and
multidimensional turbolence during collapse e.g., \cite{RSO18}.

The transition from WD to NS remnants is continuous and exhibits overlapping,
in the sense that NS masses occasionally fall below Chandrasekhar mass e.g.,
\cite{OzF16}.    Conversely, the transition from NS to BH remnants shows a
desert within the mass range, $2\appleq M/m_\odot\appleq4$ e.g., \cite{faa11}.
In
fact, low-mass BH remnants exceed $4m_\odot$ e.g., \cite{GeH03} \cite{gib18}.
Concerning X-ray transients with low-mass donors, it has been argued that
understimates of the inclination, by $10^\circ$ at least, could significantly
reduce BH mass estimates, filling the gap between the low end of BH mass
distribution and the maximum theoretical NS mass \cite{kra12}.

Submassive $(2\appleq M/m_\odot\appleq4)$ BH remnants could be the product of
WD-WD, WD-NS, NS-NS mergers after
orbital decay via gravitational radiation as in the recent event GW 170817,
where the inferred mass amounts to $2.73_{-0.01}^{+0.04}m_\odot$ \cite{aba17}.
With regard to merger product, BH is largely favoured but ultramassive NS
exhibiting a millisecond spin period cannot still be excluded
e.g., \cite{aia18}.

Models of stellar evolution could occasionally yield submassive BH remnants
after significant amount of fallback from supernovae.   If otherwise, the
lower BH mass limit from computations reads
$(M_{\rm BH})_{\rm min}\approx4m_\odot$, which is related to the He-core mass
of ZAMS progenitors where $M_{\rm ZAMS}=15m_\odot$ for $Z=Z_\odot$
\cite{RSO18}.

The rarity of the above mentioned events, i.e. orbital decay in binary systems
hosting WDs and/or NSs, and BHs following fallback from supernovae, by
itself provides an explanation to the desert between NS and BH remnants.

In this view, a nontrivial question reads: ``Could ultramassive NSs, or in
general putative BH progenitors, end their life via supernova explosion
similarly to supramassive $(M\appgeq1.44m_\odot)$ WDs?''   Ultramassive NSs
and high-density
i.e. low-mass putative BH progenitors are expected to host, among others,
bottomed and/or charmed heavy baryons e.g., \cite{pra18}.   According to a
recent investigation,
the above mentioned heavy baryons can undergo quark-level fusion to release
energy \cite{KaR17}, see also \cite{Mil17}.

A desert within the mass range, $2\appleq M/m_\odot\appleq4$, would
naturally occur if NS remnants can undergo supernova explosion leaving no
remnant.  If otherwise,
different channels should be taken into considerations.   The current paper
aims to investigate what is the case.   Stellar remnants exhibiting well
determined masses
are presented in Section \ref{stere}.   Considerations about NSs are made in
Section \ref{neusa}, where attention is focused on a density parameter related
to neutron mean density.   Considerations about BHs are made in Section
\ref{blaho}, where attention is focused on spin parameter, description in
terms of homogeneous, infinitely thin, rigidly rotating disks, and comparison
between non rotating and maximally rotating BH and NS/QS configurations.   An
upper limit to the energy released via quark-level fusion in heavy baryons is
estimated in Section \ref{quafu}.   The discussion is performed in Section
\ref{disc}.   The conclusion is presented in Section \ref{conc}.

\section{Stellar remnants}
\label{stere}

Stellar remnants are the final product of stellar evolution, where ``star''
means some kind of nucleosynthesis has occurred, to an extent depending on
ZAMS progenitor mass.

Low-mass $(0.012\appleq M/m_\odot\appleq0.06)$ BDs can perform D-burning,
while high-mass $(0.06\appleq M/m_\odot\appleq0.07)$ BDs, in addition, 
Li-burning.   After D and Li exhaustion, an inert BD can be conceived as a
remnant in the aforementioned sense.   For further details, an interested
reader is addressed to review papers e.g., \cite{Bur08} \cite{bua01}
or research articles e.g., \cite{bua97} \cite{bua98} \cite{SBM11}
\cite{boa13} \cite{Bur18}.

With regard to formation via gravitational instability vs core accretion, BD
lower mass limit is reduced to about $10M_{\rm J}=0.009546m_\odot$ regardless
of D-burning
e.g., \cite{Sch18}.   Accordingly, the above mentioned value shall be assumed
as a threshold between BDs and giant planets.   Keeping in mind ultramassive
BDs could exist \cite{FoL18}, BD upper mass limit shall be assumed as
coinciding with H-burning limit, $M_{\rm H}=0.071m_\odot$ e.g.,
\cite{Bur08} \cite{FoL18}.

Low-mass $(0.07\appleq M/m_\odot\appleq8)$ main sequence (MS) stars can
synthesize H, He, metals up to Mn, according to related ZAMS masses, ending
their life as WDs.   For nonrotating configurations where Chandrasekhar mass
is exceeded, the electron Fermi
pressure is no longer able to sustain WDs, which undergo supernova explosions
of type Ia.  Ultramassive WDs could be preserved by fast differential rotation
e.g., \cite{OBL66} \cite{OsB68} or strong magnetic fields e.g., \cite{KuM12}
\cite{DaM14}.

Large-mass $(8\appleq M/m_\odot\appleq120)$ MS stars can synthesize all
elements up to Fe, ending their life as either NS after core-collapse and
supernova explosion of type II, or BH.   For sufficiently low masses, core
collapse can be halted by Fermi pressure and the result is a NS. If otherwise,
gravitational collapse is inescapable yielding a BH.

According to recent investigations, the final fate of massive
$(M\appgeq15m_\odot)$ star is decided by different processes, such as mass
loss and onset of URCA reactions.   The result is NS and BH co-existence or
mutual exclusion in different mass ranges.   With regard to solar
metallicities, the following NS fraction for assigned ZAMS progenitor mass
range has been inferred from numerical simulations \cite{RSO18}:
\begin{lefteqnarray}
\label{eq:XNS}
&& X_{\rm NS}=\cases{
1    ~~; & $10\le M_{\rm ZAMS}/m_\odot\le15$  ; \cr
0.6  ~~; & $15\le M_{\rm ZAMS}/m_\odot\le21$  ; \cr
0    ~~; & $21\le M_{\rm ZAMS}/m_\odot\le25$  ; \cr
0.75 ~~; & $25\le M_{\rm ZAMS}/m_\odot\le28$  ; \cr
0    ~~; & $28\le M_{\rm ZAMS}/m_\odot\le60$  ; \cr
0.4  ~~; & $60\le M_{\rm ZAMS}/m_\odot\le120$ ; \cr
}
\end{lefteqnarray}
where related BH fraction is $X_{\rm BH}=1-X_{\rm NS}$.

Leaving aside fast rotation and strong magnetic fields, NS maximum mass is
slightly larger than $2m_\odot$ e.g., \cite{ASB18} \cite{RMW18}.   Conversely,
BH masses
from stellar implosions are bounded by He-core and final pre-supernova masses
of the progenitor, according to the fraction of stellar envelope which is
ejected during or prior to implosion.   Related masses depend on the ZAMS
progenitor mass as \cite{RSO18}:
\begin{lefteqnarray}
\label{eq:mBHi}
&& y=-2.049+0.4140x~~;\qquad y=M_{\rm BH,\,core}~~;\qquad x=M_{\rm ZAMS}~~; \\
\label{eq:mBHs}
&& y=15.52-0.3294x-0.02121x^2+0.003120x^3~~; \nonumber \\
&& y=M_{\rm BH,\,all}~~;\qquad x=M_{\rm ZAMS}-25.97~~;
\end{lefteqnarray}
for $15\le M_{\rm ZAMS}/m_\odot\le40$, where envelope mass of pre-supernova
progenitors cannot be neglected with respect to He-core mass, and:
\begin{lefteqnarray}
\label{eq:mBH}
&& y=5.697+7.8598\cdot10^8x^{-4.858}~~;\qquad y=M_{\rm BH,\,core}~~;\qquad
x=M_{\rm ZAMS}~~;\qquad
\end{lefteqnarray}
for $45\le M_{\rm ZAMS}/m_\odot\le120$, where envelope mass of pre-supernova
progenitors can safely be neglected with respect to He-core mass.
Occasionally, subHe-core (i.e. below He-core mass; in particular, submassive)
BH remnants can be formed from massive progenitors via fallback from
supernovae \cite{RSO18}.

\begin{figure*}[t]  
\begin{center}      
\includegraphics[scale=0.8]{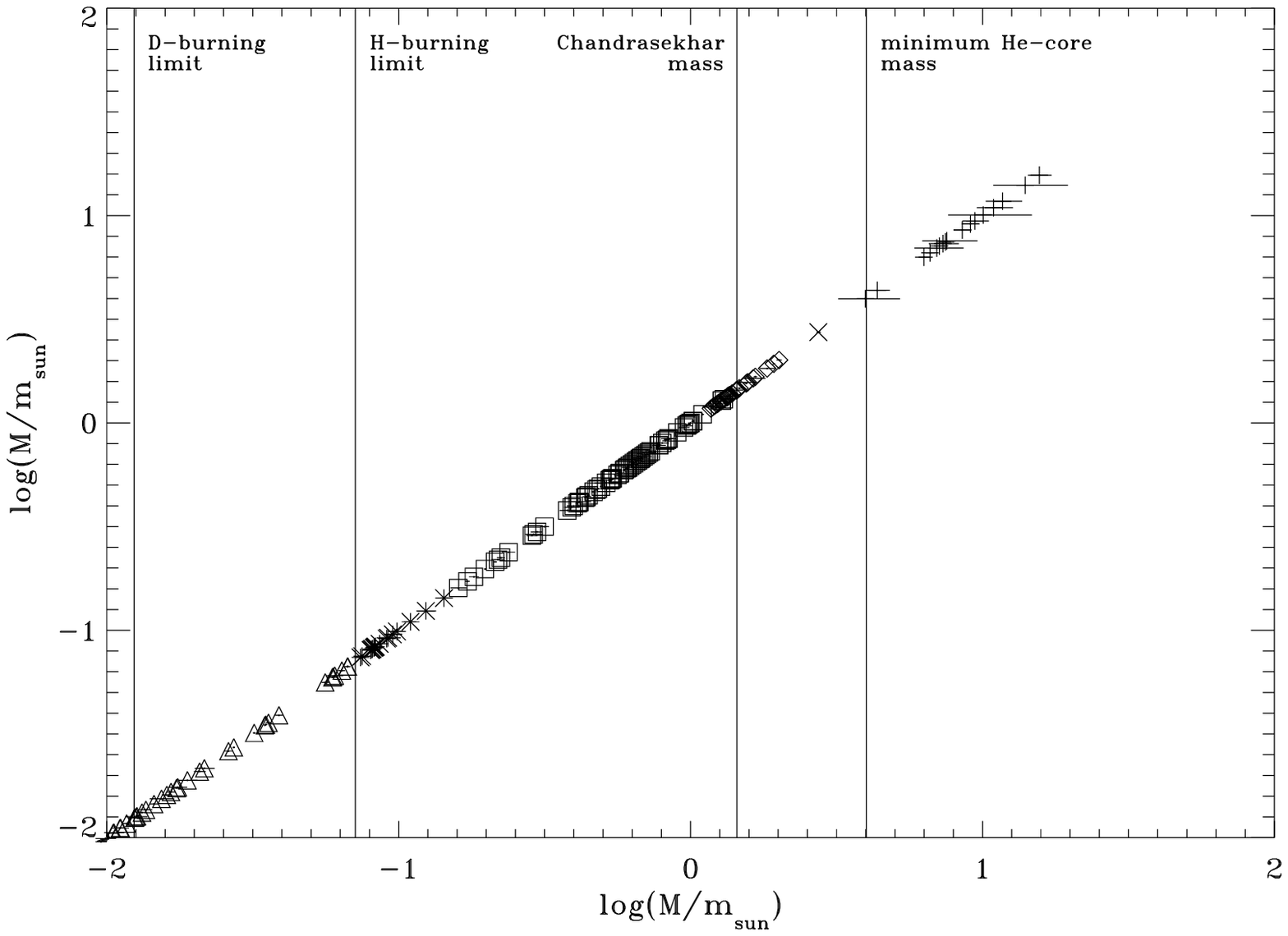}                      
\caption[ddbb]{Brown dwarf (triangles), main sequence-dwarf (asterisks), white
dwarf (squares), neutron star (diamonds), black hole (crosses) mass range for
stars listed in Tables \ref{t:mabda}-\ref{t:mabdb}, \ref{t:mams},
\ref{t:mawda}-\ref{t:mawdc},
\ref{t:mans}, \ref{t:mabh}, respectively, shown as a logarithmic
mass-to-mass relation in solar units.   The post-merger product from
the double neutron star binary GW 170817 is shown as a saltire.   To avoid
confusion, errors
are represented as horizontal bars instead of squares.   D-burning limit
($13 M_{\rm J}=0.01241 m_\odot$), H-burning limit ($0.071 m_\odot$), 
Chandrasekhar mass ($1.44\, m_\odot$), and minimum He-core mass ($4m_\odot$)
of BH progenitors are marked as vertical lines.}
\label{f:lico}     
\end{center}       
\end{figure*}

Stellar remnants with well determined masses i.e. percent error
$\Delta^\%M=100(\Delta M/M)<10$ or $\Delta M/M<0.1$ are shown in
Fig.\,\ref{f:lico} as $\log M$ vs $\log M$.
Owing to the paucity of known BH remnants at present, data affected by larger
percent error are also included.   To avoid confusion, errors are represented
as horizontal bars instead of squares.   BD, WD, NS, BH points are shown as
triangles, squares, diamonds, crosses, respectively.   Related values are
listed in Appendix \ref{a:remma}.   The post-merger product
from the double NS binary GW 170817 is shown as a saltire.   D-burning limit,
assumed to occur at $13M_{\rm J}=0.01241m_\odot$; H-burning limit, assumed to
occur at $0.071m_\odot$; Chandrasekhar mass, assumed to be $1.44m_\odot$;
minimum He-core mass of BH progenitors, inferred from numerical simulations as
$4m_\odot$ \cite{RSO18}; are marked as vertical lines.

An inspection of Fig.\,\ref{f:lico} discloses a continuous transition in mass
passing from WDs to NSs, where the two domains partially overlap.   The gap
between BDs and WDs, $0.07\appleq M/m_\odot\appleq0.16$, implies related WD
progenitors are still alive as MS dwarfs, also plotted as asterisks and listed
in Appendix \ref{a:remma}.   The gap between NSs and BHs could be interpreted
as due to selection effects and/or lack of data for massive $(M\ge15m_\odot)$
ZAMS progenitors, yielding He-core masses above $4m_\odot$ \cite{RSO18}.
Conversely the mass range, $2\appleq M/m_\odot\appleq4$, needs alternative
explanations.

To this respect, a viable possibility could be compact remnant i.e. WD-WD,
WD-NS, NS-NS, merging as recently observed for the double NS pair GW 170817,
where the total mass amounts to $2.73_{-0.01}^{+0.04}m_\odot$ including
ejected matter \cite{aba17} \cite{aia18}.   Accordingly, the merger product is
an isolated BH in absence of additional system memberships, which could
explain the presence of a desert in the above mentioned mass
range, keeping in mind isolated BHs are difficult to be detected.

On the other hand, binary systems hosting a NS and a donor MS star could yield
a BH via mass accretion onto NS and subsequent instability, where the presence
of a BH should be inferred as mass accretion goes on.   But low-mass BHs in
binary systems have not been detected up today.   Then a legitimate question
is: ``Could unstable NSs undergo supernova explosions leaving no remnant?''

\section{About neutron stars}
\label{neusa}

NS masses typically lie within the range, $1\appleq M/m_\odot\appleq2$, as can
be recognized from Fig.\,\ref{f:lico} and Table \ref{t:mans}.   NS radii
typically lie within the range, $9\appleq R/{\rm km}\appleq12$, as can be
inferred from the $M$-$R$ relation assuming reliable EOSs e.g.,
\cite{OzF16}.    Then related mean density is bounded as:
\begin{leftsubeqnarray}
\slabel{eq:rhoa}
&& \overline\rho_{\rm min}\appleq\overline\rho\appleq\overline\rho_{\rm max}
~~; \\
\slabel{eq:rhob}
&& \overline\rho_{\rm min}=\frac{m_\odot}{4\pi(12\,{\rm km})^3/3}=2.747288
\cdot10^{17}\frac{\rm kg}{\rm m^3}~~; \\
\slabel{eq:rhoc}
&& \overline\rho_{\rm max}=\frac{2m_\odot}{4\pi
(9\,{\rm km})^3/3}=1.302418\cdot10^{18}\frac{\rm kg}{\rm m^3}~~;
\label{seq:rho}
\end{leftsubeqnarray}
to be compared with nuclear saturation density, $\rho_{\rm sat}\approx2.8
\cdot10^{17}\,{\rm kg/m^3}$ e.g., \cite{OzF16} \cite{haa16}.

With regard to neutrons in atomic nuclei (intended in ordinary conditions),
mass and radius shall be assumed as $m_{\rm n}=1.675\cdot10^{-27}$\,kg and
$r_{\rm n}=10^{-15}$\,m, respectively, yielding the following neutron mean
density:
\begin{lefteqnarray}
\label{eq:rhn}
&& \overline\rho_{\rm n}=\frac3{4\pi}\frac{m_{\rm n}}{r_{\rm n}^3}=3.9988\cdot
10^{17}\frac{\rm kg}{\rm m^3}~~;
\end{lefteqnarray}
which is comparable to NS mean density and, for this reason, shall be taken as
a reference density.   On the other hand, a recent investigation has disclosed
central pressure inside protons is larger than inside NSs \cite{BEG18}.

The ratio:
\begin{lefteqnarray}
\label{eq:zin}
&& \zeta_{\rm n}=\frac{\overline\rho_{\rm n}}{\overline\rho}=\frac{m_{\rm n}}M
\frac{R^3}{r_{\rm n}^3}~~;
\end{lefteqnarray}
may be intended as a density factor, where $\zeta_n>1$ implies lower NS mean
density with respect to neutron and the contrary holds for $\zeta_{\rm n}<1$.

In explicit form, Eq.\,(\ref{eq:zin}) reads:
\begin{lefteqnarray}
\label{eq:zsk}
&& \zeta_{\rm n}=
\frac{m_{\rm n}}
{m_\odot}\frac{m_\odot}M\frac{R^3}{\rm km^3}\frac{\rm km^3}{r_{\rm n}^3}=
0.8423223\frac{(R/{\rm km})^3}{M/m_\odot}~~;
\end{lefteqnarray}
where mass and radius are expressed in solar masses and kilometers,
respectively.

\begin{figure*}[t]  
\begin{center}      
\includegraphics[scale=0.8]{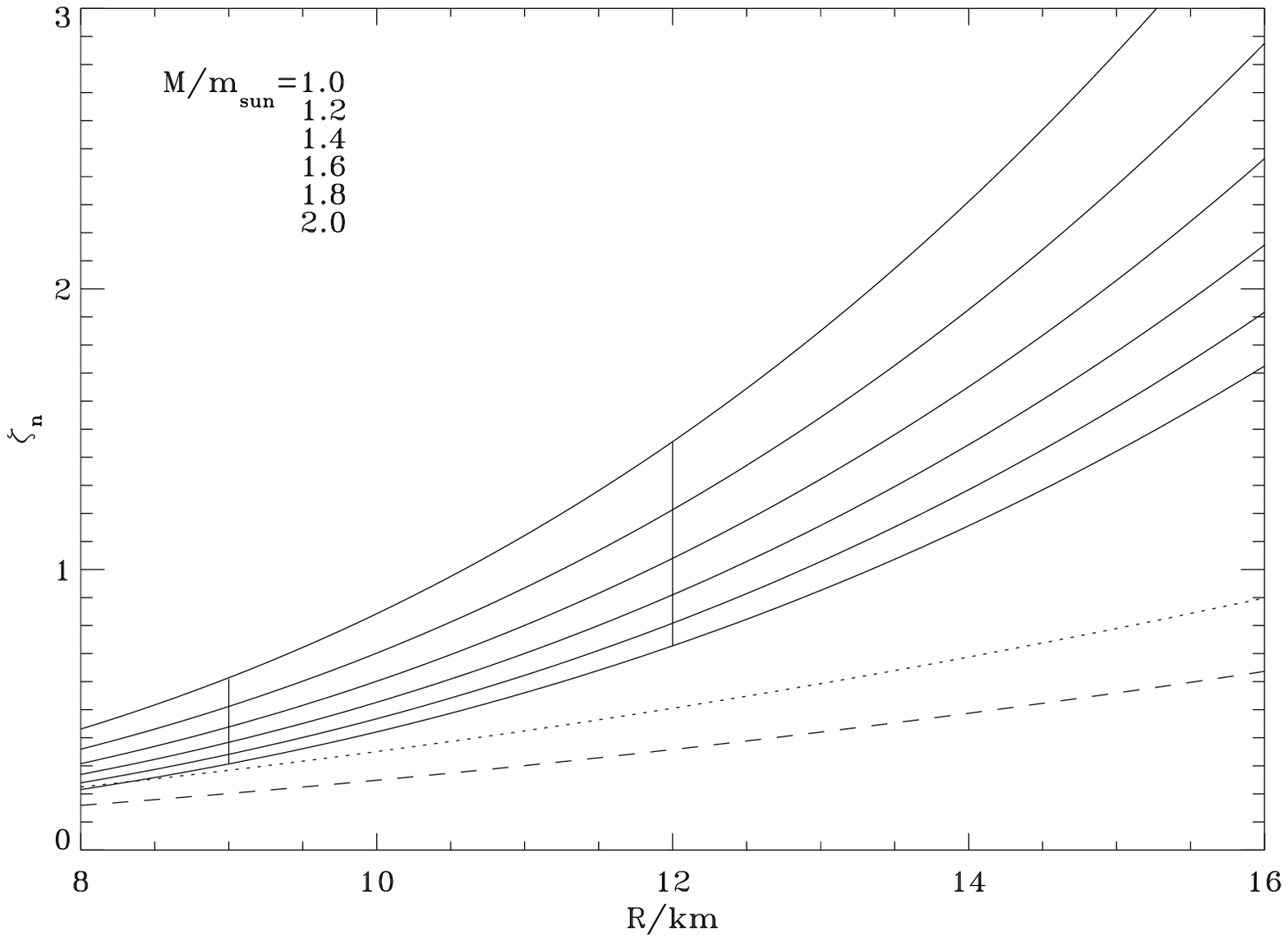}                      
\caption[ddbb]{Neutron star density factor, $\zeta_n$, vs radius, $R$/km, for
assigned mass, $M/m_\odot$.   Most of currently available or inferred data lie
within a region bounded by $R/{\rm km}=9,\,12$, and $M/m_\odot=1,\,2$. Onset 
of inescapable gravitational collapse and casual relationship is marked as a
dashed and dotted line, respectively.}
\label{f:fari}     
\end{center}       
\end{figure*}

NS density factor, $\zeta_n$, vs radius, $R$/km, for assigned mass,
$M/m_\odot$, is plotted in Fig.\,\ref{f:fari}, where most of currently
available or inferred data e.g., \cite{OzF16} lie within a region bounded by
$R/{\rm km}=9,\,12$, and $M/m_\odot=1,\,2$, e.g., \cite{OzF16}, which implies
$0.3\appleq\zeta_{\rm n}\appleq1.5$.

\begin{figure*}[t]  
\begin{center}      
\includegraphics[scale=0.8]{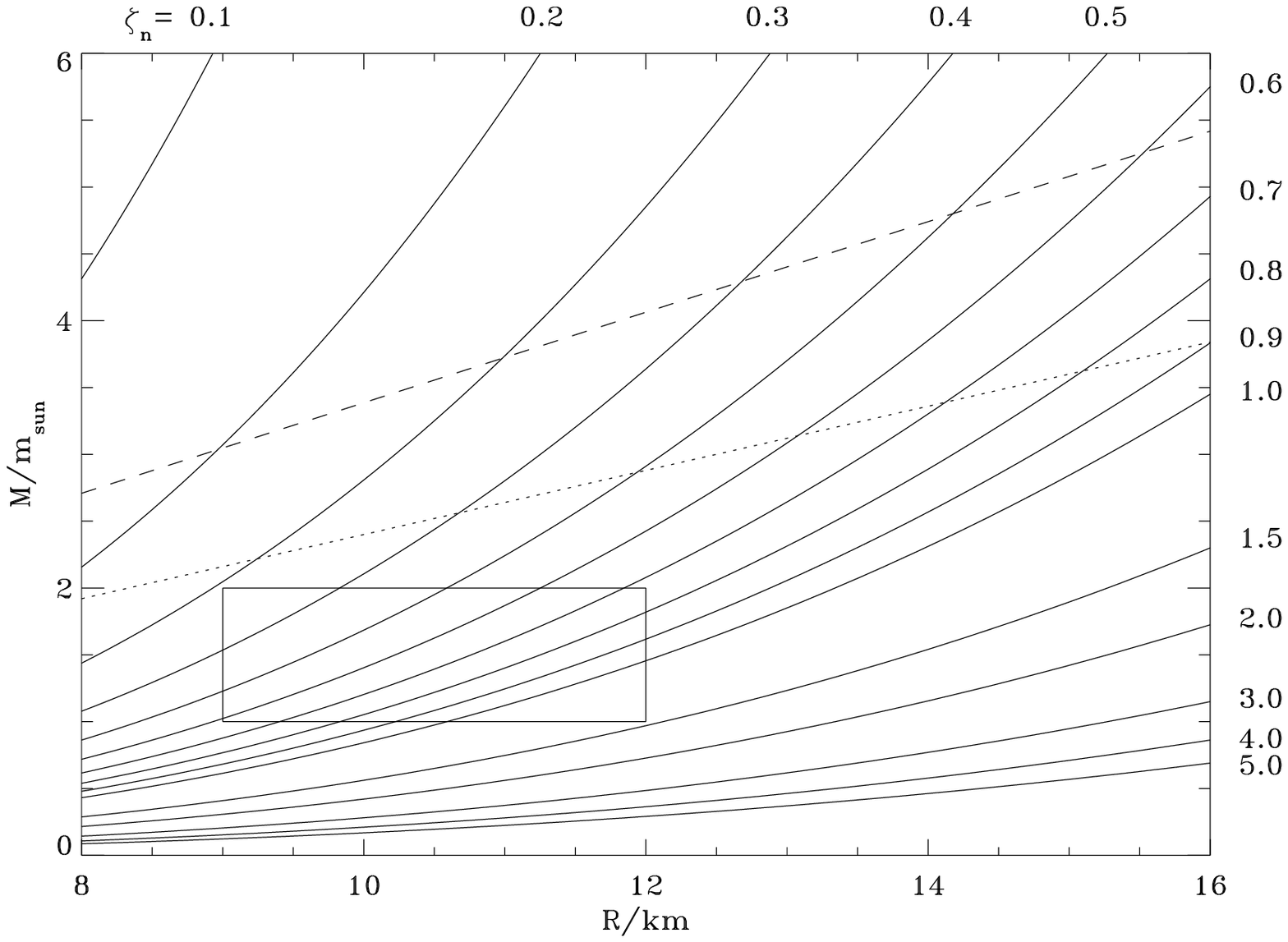}                      
\caption[ddbb]{Neutron star mass, $M/m_\odot$, vs radius, $R$/km, for assigned
density factor, $\zeta_{\rm n}$.  Most of currently available or inferred data
lie within a region bounded by $R/{\rm km}=9,\,12$, and $M/m_\odot=1,\,2$.
Onset of inescapable gravitational collapse and casual relationship is marked
as a dashed and dotted line, respectively.}
\label{f:mara}     
\end{center}       
\end{figure*}

NS mass, $M/m_\odot$, vs radius, $R$/km, for assigned density factor,
$\zeta_{\rm n}$, is plotted in Fig.\,\ref{f:mara}, where the rectangular area
is the counterpart of the region shown in Fig.\,\ref{f:fari}.

The gravitational i.e. Schwartzschild radius of a spherical-symmetric mass
distribution reads e.g., \cite{LaL66} Chap.\,XI, \S97:
\begin{lefteqnarray}
\label{eq:Rg}
&& R_{\rm g}=\frac{2GM}{c^2}~~;
\end{lefteqnarray}
and related mean density is:
\begin{lefteqnarray}
\label{eq:rhog}
&& \overline\rho_{\rm g}=\frac{3M}{4\pi R_{\rm g}^3}=\frac3{32\pi}\frac{c^6}
{G^3M^2}~~;
\end{lefteqnarray}
which scales as an inverse square mass.

Using Eq.\,(\ref{eq:Rg}), the mass as a function of the gravitational radius
reads:
\begin{lefteqnarray}
\label{eq:MRg}
&& \frac M{m_\odot}=\frac{c^2}{2Gm_\odot}\frac{R_{\rm g}}{{\rm m}}=
\frac{R_{\rm g}/{\rm m}}{(R_{\rm g})_\odot/{\rm m}}=
\frac{R_{\rm g}/{\rm km}}{(R_{\rm g})_\odot/{\rm km}}=
0.338604\frac{R_{\rm g}}{\rm km}~~;
\end{lefteqnarray}
which is plotted in Figs.\,\ref{f:fari} and \ref{f:mara} as a dashed line.

The gravitational radius, $R_{\rm g}$, via Eqs.\,(\ref{eq:zsk}) and
(\ref{eq:Rg}) can be expressed in terms of related
density parameter, $(\zeta_{\rm n})_{\rm g}$, as:
\begin{lefteqnarray}
\label{eq:Rgzn}
&& \frac{R_{\rm g}}{\rm km}=\left[\frac{c^2}{2Gm_{\rm n}}
(\zeta_{\rm n})_{\rm g}r_{\rm n}^3\right]^{1/2}\frac1{\rm km}=
20.0497[(\zeta_{\rm n})_{\rm g}]^{1/2}~~;
\end{lefteqnarray}
and related mass reads:
\begin{lefteqnarray}
\label{eq:MRgz}
&& \frac M{m_\odot}=\frac{c^2R_{\rm g}}{2Gm_\odot}=\frac{c^2}{2Gm_\odot}
\left[\frac{c^2(\zeta_{\rm n})_{\rm g}r_{\rm n}^3}{2Gm_{\rm n}}\right]^{1/2}=
6.7889[(\zeta_{\rm n})_{\rm g}]^{1/2}~~;
\end{lefteqnarray}
accordingly, $(R_{\rm g}/{\rm km},M/m_\odot)=
\{20.0497[(\zeta_{\rm n})_{\rm g}]^{1/2},
6.7889[(\zeta_{\rm n})_{\rm g}]^{1/2}\}$ are NS gravitational radius and mass
for assigned density factor, $(\zeta_{\rm n})_{\rm g}$.

Casual relationship may safely be expressed as e.g., \cite{LaP04}
\cite{ROP16}:
\begin{lefteqnarray}
\label{eq:care}
&& \frac M{m_\odot}=0.24\frac R{\rm km}~~;
\end{lefteqnarray}
which is plotted in Figs.\,\ref{f:fari} and \ref{f:mara} as a dotted line.
Larger masses at fixed radius imply causality violation.

Let $M$ and $\overline\rho$ be mass and mean density, respectively, of a
generic body.   The radius, $R_{\rm s}$, and the free-fall time,
$(t_{\rm s})_{\rm ff}$, of a dust sphere of equal mass and mean density read
e.g., \cite{AnC90}:
\begin{lefteqnarray}
\label{eq:Rs}
&& R_{\rm s}=\left(\frac3{4\pi}\frac M{\overline\rho}\right)^{1/3}~~; \\
\label{eq:taus}
&& (t_{\rm s})_{\rm ff}=\left(\frac{3\pi}{32}\frac1{G\overline\rho}\right)^
{1/2}~~;
\end{lefteqnarray}
respectively.

The particularization to $\overline\rho=\overline\rho_g$ and the
combination with Eqs.\,(\ref{eq:Rg}) and (\ref{eq:rhog}) yields after
little algebra:
\begin{lefteqnarray}
\label{eq:tfg}
&& (t_{\rm g})_{\rm ff}=\left(\frac{3\pi}{32}\frac1{G\overline\rho_{\rm g}}
\right)^{1/2}=\frac\pi2\frac{R_{\rm g}}c~~;
\end{lefteqnarray}
which reads $(t_{\rm g})_{\rm ff}\approx0.015$\,ms for the sun.

The combination of Eqs.\,(\ref{eq:Rg}), (\ref{eq:Rs}), (\ref{eq:taus}), after
some algebra yields:
\begin{lefteqnarray}
\label{eq:RsRg}
&& \frac{R_{\rm s}}{R_{\rm g}}=\left(\frac{M_0^2}{M^2}\frac{\overline\rho_0}
{\overline\rho}\right)^{1/3}~~;\qquad M_0^2\overline\rho_0=\frac3{32\pi}
\frac{c^6}{G^3}~~;
\end{lefteqnarray}
which scales as $M^{-2/3}\overline\rho^{-1/3}$.

For fixed $\overline\rho$, Eq.\,(\ref{eq:RsRg}) reduces to:
\begin{lefteqnarray}
\label{eq:RM0}
&& \frac{R_{\rm s}}{R_{\rm g}}=\left(\frac{M_0}{M}\right)^{2/3}~~;\qquad
M_0=\frac{c^3(t_{\rm s})_{\rm ff}}{\pi G}~~;
\end{lefteqnarray}
which scales as $M^{-2/3}$.

For fixed $M$, Eq.\,(\ref{eq:RsRg}) reduces to:
\begin{lefteqnarray}
\label{eq:RM0}
&& \frac{R_{\rm s}}{R_{\rm g}}=\left(\frac{\overline\rho_0}{\overline\rho}
\right)^{1/3}~~;\qquad \overline\rho_0=\frac3{32\pi}\frac{c^6}{G^3M^2}~~;
\end{lefteqnarray}
which scales as $\overline\rho^{-1/3}$.

\begin{figure*}[t]  
\begin{center}      
\includegraphics[scale=0.8]{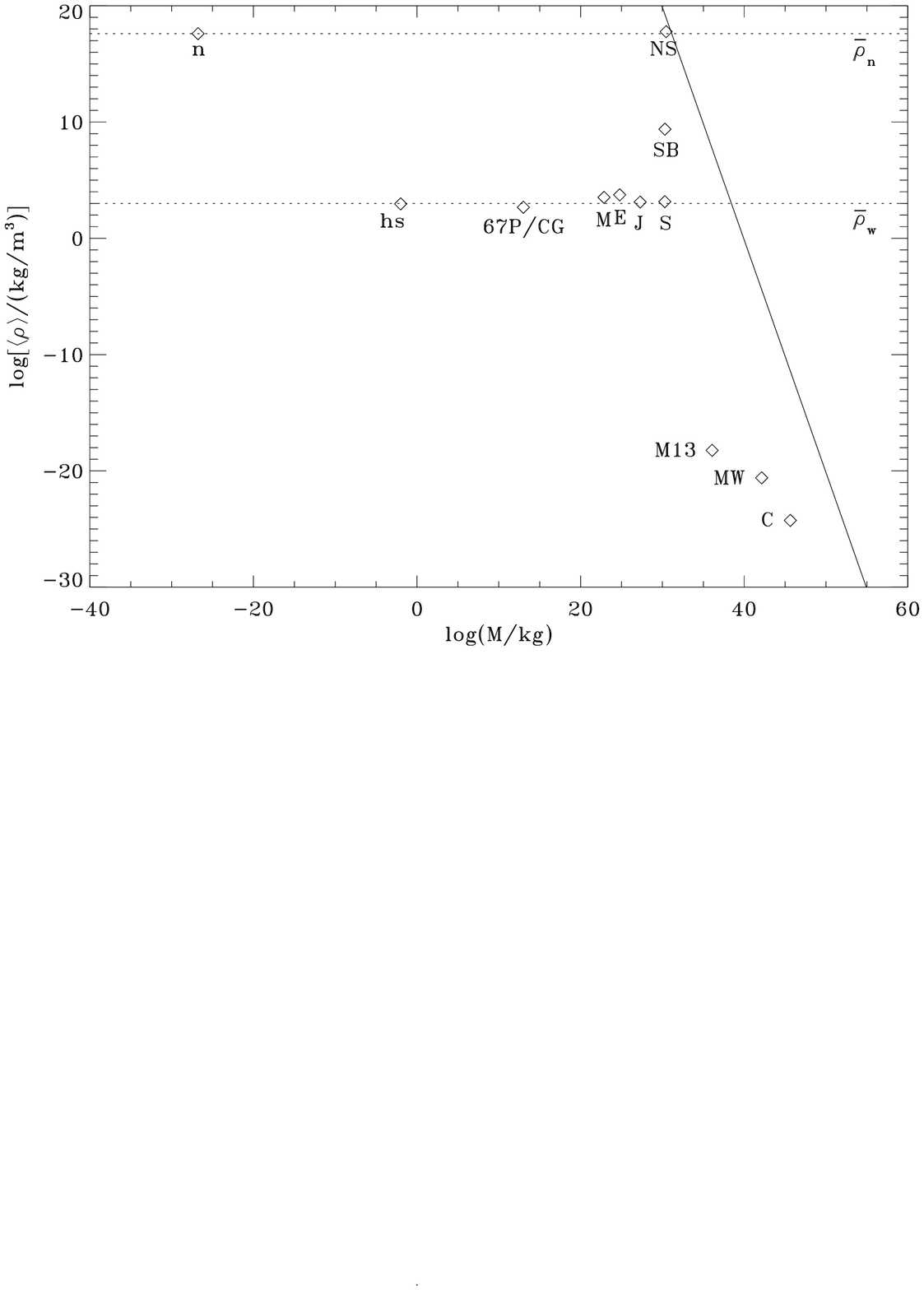}                      
\caption[ddbb]{Mean density vs mass in the logarithmic plane for the following
bodies: neutron (n); hail bead (hb); comet (67P/Churyumov-Gerasimenko,
67P/CG); Moon (M); Earth (E); Jupiter (J); Sun (S); white dwarf (Syrius B,
SB); neutron star (PSR J0337+1715, assumed radius, NS); globular cluster (M13,
M13); galaxy (Milky Way, MW); galaxy cluster (Coma, C).   The threshold of
inescapable gravitational collapse, $R_{\rm s}=R_{\rm g}$, is shown as a full
line. Neutron mean density and water mean density are marked as dotted
horizontal lines.}
\label{f:cost}     
\end{center}       
\end{figure*}

Turning to the general case, the threshold of inescapable gravitational
collapse is $R_{\rm s}=R_{\rm g}$ and Eq.\,(\ref{eq:RsRg}) reduces to
$M^2\overline\rho=M_0^2\overline\rho_0$ or:
\begin{lefteqnarray}
\label{eq:lRM}
&& \log\frac{\overline\rho}{\rm kg/m^3}=\log\left(\frac3{32\pi}\frac{c^6}{G^3}
\frac{\rm m^3}{\rm kg^3}\right)-2\log\frac M{\rm kg}~~;
\end{lefteqnarray}
which is shown in Fig.\,\ref{f:cost} as a full line.
Dotted horizontal lines mark neutron mean density,
$\overline\rho_{\rm n}=3.9988\cdot10^{17}\,{\rm kg/m^3}$, and water mean
density, $\overline\rho_{\rm w}=0.9998\cdot10^3\,{\rm kg/m^3}$, respectively.

For comparison, the following bodies are also plotted in Fig.\,\ref{f:cost} as
diamonds: neutron (n); hail bead (hb); comet (67P/Churyumov-Gerasimenko,
67P/CG); Moon (M); Earth (E); Jupiter (J); Sun (S); white dwarf (Syrius B,
SB); neutron star (PSR J0337+1715, assumed radius, NS); globular cluster (M13,
M13); galaxy (Milky Way, MW); galaxy cluster (Coma, C).
Related data are listed in Table \ref{t:cost}.
\begin{table*}
\caption[par]{Mass, $M$; mean density, $\overline\rho$; radius of associated
sphere, $R_{\rm s}$; normalized radius of associated sphere with respect to
gravitational radius of associated sphere, $R_{\rm s}/R_{\rm g}$; of selected
bodies ranging from micro
to macro cosmos.   Symbol caption: n - neutron; hb - hail bean; 67P/CG - comet
(67P/Churyumov-Gerasimenko); M - Moon; E - Earth; J - Jupiter; S - Sun; SB - 
white dwarf (Syrius B); NS - neutron star (PSR J0337+1715, assumed radius);
M13
- globular cluster (M13); MW - galaxy (Milky Way); C - galaxy cluster (Coma).}
\label{t:cost}
\begin{center}
\begin{tabular}{lllll}
\hline
\multicolumn{1}{c}{symbol\phantom{oo}}         &
\multicolumn{1}{c}{$M$/kg}                     &
\multicolumn{1}{c}{$\overline\rho$/(kg/m$^3$)} &
\multicolumn{1}{c}{$R_{\rm s}$/m}              &
\multicolumn{1}{c}{$R_{\rm s}/R_{\rm g}$}      \\
\hline\noalign{\smallskip}
n      & 1.6749E$-$27 & 3.9988E$+$17 & 1.0000E$-$15 & 4.8729E$+$60 \\
hb     & 1.0000E$-$02 & 9.1700E$+$02 & 1.3757E$-$02 & 1.1228E$+$49 \\
67P/CG & 1.0000E$+$13 & 4.7000E$+$02 & 1.7190E$+$03 & 1.1574E$+$17 \\
M      & 7.3420E$+$22 & 3.3462E$+$03 & 1.7367E$+$06 & 1.5927E$+$10 \\
E      & 5.9726E$+$24 & 5.5140E$+$03 & 6.3709E$+$06 & 7.1824E$+$08 \\
J      & 1.8986E$+$27 & 1.3260E$+$03 & 6.9920E$+$07 & 2.4797E$+$07 \\
S      & 1.9886E$+$30 & 1.4112E$+$03 & 6.9548E$+$08 & 2.3549E$+$05 \\
SB     & 2.0243E$+$30 & 2.4139E$+$09 & 5.8500E$+$06 & 1.9458E$+$03 \\
NS     & 2.8591E$+$30 & 5.8963E$+$17 & 1.0500E$+$04 & 2.4728E$+$00 \\
M13    & 1.1931E$+$36 & 5.8469E$-$19 & 7.8685E$+$17 & 4.4405E$+$08 \\
MW     & 1.3562E$+$42 & 2.5446E$-$21 & 5.0297E$+$20 & 2.4972E$+$05 \\
C      & 4.3549E$+$45 & 5.5292E$-$25 & 1.2343E$+$23 & 1.9083E$+$04 \\
\noalign{\smallskip}
\hline                             
\end{tabular}                      
\end{center}                       
\end{table*}                       

An inspection of Fig.\,\ref{f:cost} discloses NSs (including QSs and
hybrid stars) are systems (where laws of ordinary physics hold) closest to 
BHs (where laws of ordinary physics no longer hold).

\section{About black holes}
\label{blaho}

BH progenitors are characterized by inescapable gravitational collapse where,
from the standpoint of general relativity, the final configuration may be
conceived as a mass point, or spindle, or flat disk.   Related observables are
mass, angular momentum, and electric charge, leaving aside thermodynamics.
The situation is far more
complex in the framework of superstring theories, where 7 additional
dimensions are taken into consideration, and supersymmetric (SUSY) theories,
where gravitational, electromagnetic, weak and strong interaction can be
unified.   For further details, an interested reader is addressed to specific
textbooks e.g., \cite{Gre99}.   In any case, BHs may safely be thought of as 
single entities, according to the following considerations.

Particles lose their identity within BHs, in the sense their self energy is
negligible with respect to tidal energy from BH.   Matter therein fills so
tiny volume that extra dimensions (if any) are comparable to usual dimensions,
implying BHs attain a pre-big bang state.   Then the four known interactions
therein
are expected to be unified into a single interaction i.e. supergravity.
Outside event horizon, supergravity necessarily reduces to ordinary gravity
and the four usual dimensions dominate on the hidden remaining seven.

Rotating BHs in Kerr metric are characterized by spin parameter e.g.,
\cite{Tho74} \cite{pia18} as:
\begin{lefteqnarray}
\label{eq:aBH}
&& a_{\rm BH}^*=\frac{cJ_{\rm BH}}{GM_{\rm BH}^2}=\frac{c^2}{2GM_{\rm BH}}
\frac{2J_{\rm BH}}{cM_{\rm BH}}=\frac{2J_{\rm BH}}{cM_{\rm BH}R_{\rm g}}~~;
\quad0\le a_{\rm BH}^*\le1~~;
\end{lefteqnarray}
accordingly, BH angular momentum reads:
\begin{lefteqnarray}
\label{eq:JBH}
&& J_{\rm BH}=\frac12M_{\rm BH}a_{\rm BH}^*cR_{\rm g}~~.
\end{lefteqnarray}

In classical mechanics, the angular momentum of a rigidly rotating (in
particular, infinitely thin) homogeneous disk is:
\begin{lefteqnarray}
\label{eq:Jd}
&& J_{\rm d}=\frac12M_{\rm d}v_{\rm d}R_{\rm d}~~;
\end{lefteqnarray}
where $v_{\rm d}=\Omega_{\rm d}R_{\rm d}$ is disk equatorial velocity and
$T_{\rm d}=2\pi/\Omega_{\rm d}=2\pi R_{\rm d}/v_{\rm d}$ is disk period.

A principle of corresponding states can be established via
Eqs.\,(\ref{eq:JBH}) and (\ref{eq:Jd}) between spinning BHs and rigidly
rotating, infinitely thin, homogeneous disks.
\begin{trivlist}%
\item[\hspace\labelsep{\bf }] \sl
Spinning BHs in Kerr metrics can be related to rigidly rotating, infinitely
thin, homogeneous disks in classical mechanics, of equal mass,
$M_{\rm d}=M_{\rm BH}$, radius equal to
gravitational radius, $R_{\rm d}=R_{\rm g}$, and equatorial velocity,
$v_{\rm d}=a_{\rm BH}^*c$, $0\le a_{\rm BH}^*\le1$.
\end{trivlist}
In this view, BH period and moment of inertia read:
\begin{lefteqnarray}
\label{eq:TBH}
&& T_{\rm BH}=\frac{2\pi}{\Omega_{\rm BH}}=\frac{2\pi}
{a_{\rm BH}^*c/R_{\rm g}}=\frac{2\pi}{a_{\rm BH}^*c}\frac{2GM_{\rm BH}}{c^2}
=\frac{4\pi GM_{\rm BH}}{a_{\rm BH}^*c^3}~~; \\
\label{eq:IBH}
&& I_{\rm BH}=\frac12M_{\rm BH}R_{\rm g}^2=\frac12M_{\rm BH}\left(\frac{2G
M_{\rm BH}}{c^2}\right)^2=\frac{2G^2M_{\rm BH}^3}{c^4}~~;
\end{lefteqnarray}
which allow comparison with NS/QS counterparts.

To this respect, it is worth remembering descriptions in general relativity
can be performed, in some special cases, using classical concepts.   For
instance, a logarithmic gravitational potential, together with an allowance
for space-time curvature, provide laws of motion for a free particle near a
nonrotating BH, which coincide with their counterparts formulated in general
relativity \cite{ShL18}.

With regard to stable maximum mass configurations of fixed spin frequency,
calculations were performed in a recent investigation \cite{lia16} for
axisymmetric, rigidly rotating NSs and QSs, using 4 and 3 different EOSs,
respectively.

Let $M_{\rm max}$ be maximum
mass of equilibrium configurations mentioned above, and let $R_{\rm max}$,
$I_{\rm max}$, be related equatorial radius and moment of inertia,
for assigned angular velocity, $\Omega$.    Let $M_{\rm TOV}$, $R_{\rm TOV}$,
$I_{\rm TOV}$, be their counterparts in the special case of zero spin
frequency i.e. nonrotating configurations.

The dependence of $(M_{\rm max}, R_{\rm max}, I_{\rm max})$ on angular
velocity or rotation period can be approximated by the following relations
\cite{lia16}:
\begin{lefteqnarray}
\label{eq:Mmax}
&& \frac{M_{\rm max}}{m_\odot}=\frac{M_{\rm TOV}}{m_\odot}\left[1+\alpha\left(
\frac{T}{\rm ms}\right)^\beta\right]~~; \\
\label{eq:Rmax}
&& \frac{R_{\rm max}}{\rm km}=C+A\left(\frac T{\rm ms}\right)^B~~; \\
\label{eq:Imax}
&& \frac{I_{\rm max}}{\rm 10^{38}kg\,m^2}=\frac{M_{\rm max}}{m_\odot}\left(
\frac{R_{\rm max}}{\rm km}\right)^2\frac{aC_\odot/100}{1+\exp[-kT(1-q/T)]}~~; 
\end{lefteqnarray}
where $T=2\pi/\Omega$ is period, $C_\odot$ is solar mass coefficient, and
remaining parameters depend on
EOS.   Related values are listed in the parent paper \cite{lia16} Table 1 and
an interested reader is addressed therein for further details.

The following profile parameter:
\begin{lefteqnarray}
\label{eq:eta}
&&\eta=\frac I{MR_{\rm eq}^2}=\frac{I\Omega}{MR_{\rm eq}^2\Omega}=\frac J
{MR_{\rm eq}^2v_{\rm eq}/R_{\rm eq}}=\frac J{Mv_{\rm eq}R_{\rm eq}}~~;
\end{lefteqnarray}
where $R_{\rm eq}$ is equatorial radius and $J$ angular momentum assuming
rigid rotation, via Eq.\,(\ref{eq:Imax}) reads in particular:
\begin{lefteqnarray}
\label{eq:etm}
&&\eta_{\rm max}=\frac{I_{\rm max}}{M_{\rm max}R_{\rm max}^2}=\frac
{I_{\rm max}/(10^{38}{\rm kg\,m^2})10^{38}{\rm kg\,m^2}}{(M_{\rm max}/m_\odot)
m_\odot(R_{\rm max}/{\rm km})^2{\rm km^2}} \nonumber \\
&& \phantom{\eta_{\rm max}}=
\frac{aC_\odot/100}{1+\exp
[-kT_{\rm max}(1-q/T_{\rm max})]}\frac{10^{38}{\rm kg\,m^2}}
{C_\odot10^{30}{\rm kg\,10^6m^2}} \nonumber \\
&& \phantom{\eta_{\rm max}}=
\frac a{1+\exp[-kT_{\rm max}(1-q/T_{\rm max})]}~~;
\end{lefteqnarray}
with regard to maximum mass configurations for assigned angular velocity.

The special case of TOV and equatorial breakup (EQB) configuration reads:
\begin{lefteqnarray}
\label{eq:eTOV}
&& \eta_{\rm TOV}=\frac{I_{\rm TOV}}{M_{\rm TOV}R_{\rm TOV}^2}=a~~; \\
\label{eq:eEQB}
&& \eta_{\rm EQB}=\frac{I_{\rm EQB}}{M_{\rm EQB}R_{\rm EQB}^2}=
\frac a{1+\exp[-kT_{\rm EQB}(1-q/T_{\rm EQB})]}~~;
\end{lefteqnarray}
respectively, where $T_{\rm EQB}$ is the spin period at equatorial breakup.

Realistic values of $I_{\rm TOV}$ and $\eta_{\rm TOV}$ imply the parameter,
$a$, has to be divided by $100/C_\odot$ according to Eq.\,(\ref{eq:Imax}).
In fact, related values listed in the parent paper \cite{lia16}
Table 1 have to be read as $\eta_{\rm TOV}$ instead of $a$ according to
Eq.\,(\ref{eq:eTOV}).
%
%
%

The counterpart of BH spin parameter, $a_{\rm BH}^*$, defined by
Eq.\,(\ref{eq:aBH}), for NS/QS reads: 
\begin{lefteqnarray}
\label{eq:aNQ}
&& a_{\rm NQ}^*=\frac{cJ}{GM^2}~~;
\end{lefteqnarray}
which, in the case under discussion of rigid rotation, reduces to:
\begin{lefteqnarray}
\label{eq:aNQr}
&& a_{\rm NQ}^*=\frac{cI\Omega}{GM^2}=\frac{2\pi cI}{GM^2T}~~.
\end{lefteqnarray}

In convenient units, Eq.\,(\ref{eq:aNQr}) translates into:
\begin{lefteqnarray}
\label{eq:aNQu}
&& a_{\rm NQ}^*=\frac{2\pi C_{\rm c}(10^8{\rm m/s})I/(10^{38}{\rm kg\,m^2})
10^{38}{\rm kg\,m^2}}{C_{\rm G}[10^{-11}{\rm m^3/(kg\,s^2})](M/m_\odot)^2
(C_\odot10^{30}{\rm kg})^2(T/{\rm ms}){\rm 10^{-3}s}} \nonumber \\
&& \phantom{a_{\rm NQ}^*}
=\frac{2\pi C_{\rm c}}{C_{\rm G}
C_\odot^2}\frac{I/(10^{38}{\rm kg\,m^2})}{(M/m_\odot)^2(T/{\rm ms})}~~; \\
\label{eq:cons}
&& c=C_{\rm c}10^8{\rm m/s}~~;\quad G=C_{\rm G}10^{-11}{\rm m^3/(kg\,s^2})~~;
\quad m_\odot=C_\odot10^{30}{\rm kg}~~;
\end{lefteqnarray}
where $2\pi C_{\rm c}/(C_{\rm G}C_\odot^2)=0.713893$.

Parameters related to TOV and EQB configurations, for seven different EOSs,
are listed in Table \ref{t:nqrt}, where data are taken or inferred from the
parent paper \cite{lia16}.
\begin{table*}
\caption[par]{Comparison between TOV and EQB configuration parameters (first
and second line of each panel) and additional parameters of EQB configurations
(third line of each panel).   Parameters are specified on the head of the
Table as: mass, equatorial radius, moment of inertia, profile parameter (first
and second line of each panel), and: angular momentum, specific angular
momentum, spin parameter, period (third line of each panel).   EOSs related to
each panel are (from top to bottom): BCPM, BSk20, BSk21, Shen, CIDDM, CDDM1,
CDDM2, where the first four are applied to NSs and the last three to QSs.
Data are taken or inferred from the parent paper \cite{lia16}.   See text for
further details.}
\label{t:nqrt}
\begin{center}
\begin{tabular}{llll}
\hline
\multicolumn{1}{c}{$M_{\rm TOV}/m_\odot$} &
\multicolumn{1}{c}{$R_{\rm TOV}/{\rm km}$} &
\multicolumn{1}{c}{$I_{\rm TOV}/(10^{38}{\rm kg\,m^2})$} &
\multicolumn{1}{c}{$\eta_{\rm TOV}$} \\
\multicolumn{1}{c}{$M_{\rm EQB}/m_\odot$} &
\multicolumn{1}{c}{$R_{\rm EQB}/{\rm km}$} &
\multicolumn{1}{c}{$I_{\rm EQB}/(10^{38}{\rm kg\,m^2})$} &
\multicolumn{1}{c}{$\eta_{\rm EQB}$} \\
\multicolumn{1}{c}{$J_{\rm EQB}/(\rm10^{42}kg\,m^2/s)$} &
\multicolumn{1}{c}{$j_{\rm EQB}/({\rm10^{12}m^2/s})$} &
\multicolumn{1}{c}{$a_{\rm EQB}^*$} &
\multicolumn{1}{c}{$T_{\rm EQB}/{\rm ms}$} \\
\hline\noalign{\smallskip}
1.98000D+00 & 9.94100D$+$00 & 1.75445D$+$00 & 4.50900D$-$01 \\
2.33809D+00 & 1.33165D$+$01 & 2.85700D$+$00 & 3.46524D$-$01 \\
1.37494D+00 & 6.91429D$-$01 & 6.68154D$-$01 & 5.58400D$-$01 \\
\hline
2.17000D+00 & 1.01700D$+$01 & 2.10391D$+$00 & 4.71400D$-$01 \\
2.57642D+00 & 1.34191D$+$01 & 3.50300D$+$00 & 3.79699D$-$01 \\
1.58465D+00 & 7.96888D$-$01 & 6.98828D$-$01 & 5.39100D$-$01 \\
\hline
2.28000D+00 & 1.10800D$+$01 & 2.69288D$+$00 & 4.83800D$-$01 \\
2.72701D+00 & 1.47324D$+$01 & 4.36800D$+$00 & 3.71120D$-$01 \\
1.67150D+00 & 8.40564D$-$01 & 6.96425D$-$01 & 6.02100D$-$01 \\
\hline
2.18000D+00 & 1.24000D$+$01 & 2.73421D$+$00 & 4.10200D$-$01 \\
2.59936D+00 & 1.68886D$+$01 & 4.67500D$+$00 & 3.17095D$-$01 \\
1.58203D+00 & 7.95568D$-$01 & 6.91513D$-$01 & 7.14300D$-$01 \\
\hline
2.09000D+00 & 1.24300D$+$01 & 2.84658D$+$00 & 4.43300D$-$01 \\
2.92295D+00 & 1.94749D$+$01 & 8.64500D$+$00 & 3.92154D$-$01 \\
2.23196D+00 & 1.12241D$+$00 & 8.67599D$-$01 & 8.32600D$-$01 \\
\hline
2.21000D+00 & 1.39900D$+$01 & 3.65813D$+$00 & 4.25300D$-$01 \\
3.09282D+00 & 2.24108D$+$01 & 1.16700D$+$01 & 3.77805D$-$01 \\
2.38033D+00 & 1.19702D$+$00 & 8.74453D$-$01 & 9.96000D$-$01 \\
\hline
2.45000D+00 & 1.57600D$+$01 & 5.08840D$+$00 & 4.20500D$-$01 \\
3.44237D+00 & 2.51923D$+$01 & 1.63400D$+$01 & 3.76116D$-$01 \\
2.65131D+00 & 1.33329D$+$00 & 8.75097D$-$01 & 1.12490D$+$00 \\
\noalign{\smallskip}
\hline                             
\end{tabular}                      
\end{center}                       
\end{table*}                       
More specifically, different panels correspond to
different EOSs  concerning NSs (first four) and QSs (last three).   Masses,
equatorial radii, moments of inertia, profile parameters, for TOV and EQB
configurations, are listed on the first and second line of each panel.
Angular momenta, specific angular momenta, spin parameters, periods, for EQB
configurations, are listed on the third line of each panel.    As outlined in
the parent paper \cite{lia16}, the effect of rotation is increasing mass,
equatorial radius, and moment of inertia.   On the other hand, the profile
parameter, $\eta$, appears to be lowered by rotation.

The profile parameter, $\eta$, can empirically be related to the compactness
parameter, $\beta^*=(1/2)(R_{\rm g}/R)$, as:
\begin{lefteqnarray}
\label{eq:SLB}
&& \eta=(0.237\mp0.008)[1+2.844\beta^*+18.910(\beta^*)^4]~~;
\end{lefteqnarray}
for a large EOS set \cite{LaP01} \cite{SLB10} \cite{SLB13}, and:
\begin{lefteqnarray}
\label{eq:wea}
&& \eta=0.207+0.857\beta^*\mp0.011~~;
\end{lefteqnarray}
for a limited set of microscopic EOS \cite{wea18}.

Values of $(\beta^*,\eta)$ inferred for single objects are shown in Table
\ref{t:beta}, where the first row relates to PSR J0737-3039A \cite{ROq16} and
the other two to PSR J03478+0432 \cite{Zha17} for lower and upper mass limit,
respectively.   In the former case, a fiducial value of moment of inertia
has been assumed together with a consistent fiducial value of radius.
For further details, an interested reader is addressed to the parent paper
\cite{ROq16}.
\begin{table*}
\caption[par]{Mass, $M$, radius, $R$, gravitational radius,
$R_{\rm g}$, moment of inertia, $I$, compactness parameter,
$\beta^*=(1/2)(R_{\rm g}/R)$, profile parameter, $\eta=I/(MR^2)$, inferred for
PSR J0737-3039A \cite{ROq16} (first row) and PSR J03478+0432 \cite{Zha17} 
for lower and upper mass limit (second and third row, respectively).   See 
text for further details.}
\label{t:beta}
\begin{center}
\begin{tabular}{llllll}
\hline
\multicolumn{1}{c}{$M/m_\odot$}                &
\multicolumn{1}{c}{$R$/km}                     &
\multicolumn{1}{c}{$R_{\rm g}$/km}             &
\multicolumn{1}{c}{$I/({\rm 10^{38}kg\,m^2})$} &
\multicolumn{1}{c}{$\beta^*$}                  &
\multicolumn{1}{c}{$\eta$}                     \\
\hline\noalign{\smallskip}
1.338 & 13     & 3.951 & 1.5   & 0.1520 & 0.3336 \\
1.97  & 12.957 & 5.817 & 1.906 & 0.2245 & 0.2898 \\
2.05  & 12.143 & 6.053 & 1.557 & 0.2492 & 0.2557 \\
\noalign{\smallskip}
\hline                             
\end{tabular}                      
\end{center}                       
\end{table*}                       

\begin{figure*}[t]  
\begin{center}      
\includegraphics[scale=0.8]{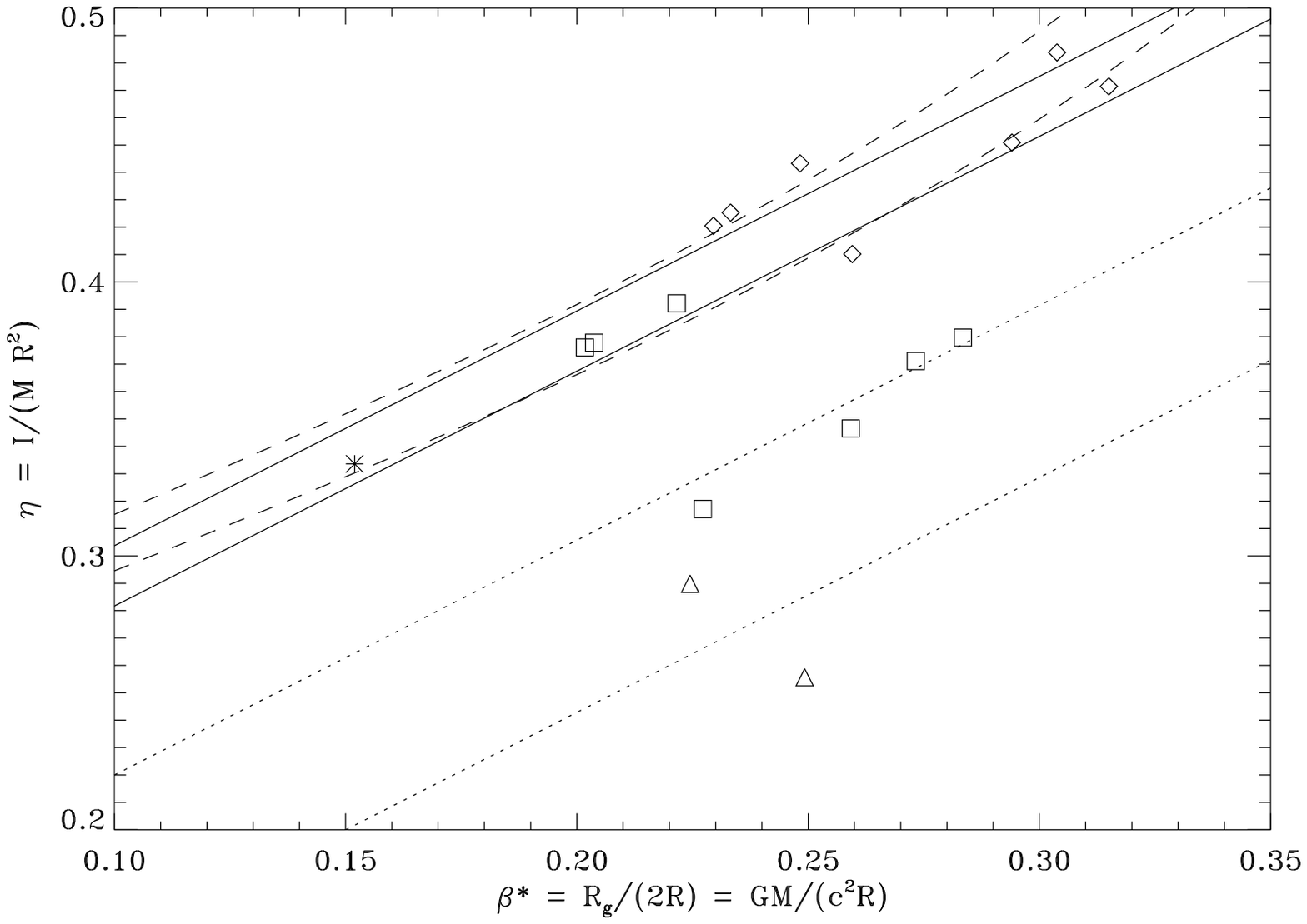}                      
\caption[ddbb]{Profile parameter, $\eta$, vs compactness parameter, $\beta^*$,
for TOV (diamonds) and EQB (squares) configurations listed in Table
\ref{t:nqrt}.   Also plotted for comparison are PSR J0737-3039A (asterisk),
inferred from fiducial and consistent values of $I$ and $R$ \cite{ROq16}, and
PSR J0348+0432 (triangles) related to lower (left) and upper
(right) mass limit \cite{Zha17}.   Empirical fits  from different EOS sets are
shown as a dashed \cite{LaP01} \cite{SLB10} \cite{SLB13} and full \cite{wea18}
band, respectively.   Dotted lines are parallel to the full band and related
intercepts read 0.1343 (selected for comparison with lower EQB configurations)
and 0.0715 [selected for fitting to BH limit, $(\beta^*,\eta)=(1/2,1/2)$].
See text for further details.}
\label{f:beta}     
\end{center}       
\end{figure*}

The compactness parameter, $\beta^*$, can be inferred from results listed in
Table \ref{t:nqrt}: related $(\beta^*,\eta)$ are plotted in
Fig.\,\ref{f:beta} as diamonds (TOV) and squares (EQB).
Empirical fits via Eqs.\,(\ref{eq:SLB}) and (\ref{eq:wea}) are shown as a
dashed and full band, respectively.   Also included for comparison are
PSR J0737-3039A (asterisk) and PSR J0348+0432 (triangles) related to lower
(left) and upper (right) mass limit.

An inspection of Fig.\,\ref{f:beta} discloses TOV configurations are
consistent or marginally inconsistent with the above mentioned fits, while EQB
configurations are consistent for QSs and largely inconsistent (but lying on a
parallel sequence) for NSs.   It is worth remembering classical and general
relativistic limit relate to $\beta^*\to0$ and $\beta^*\to1/2$, respectively.
Dotted lines parallel to the full band are shown for comparison with lower EQB
configurations (intercept arbitrarily chosen equal to 0.1343) and for fitting
BH limit, $(\beta^*,\eta)=(1/2,1/2)$ (intercept equal to 0.0715).

With regard to BHs in Kerr metric, the particularization of
Eqs.\,(\ref{eq:JBH}), (\ref{eq:TBH}), (\ref{eq:IBH}), (\ref{eq:eta}), to
nonrotating $(a_{\rm BH}^*=0)$ and maximally rotating $(a_{\rm BH}^*=1)$ BHs
yields:
\begin{lefteqnarray}
\label{eq:JBE}
&& J_{\rm EQB}=\frac12M_{\rm EQB}\,c\,(R_{\rm g})_{\rm EQB}~~; \\
\label{eq:TBE}
&& T_{\rm EQB}=\frac{2\pi(R_{\rm g})_{\rm EQB}}c~~; \\
\label{eq:IgT}
&& I_{\rm TOV}=\frac12M_{\rm TOV}[(R_{\rm g})_{\rm TOV}]^2~~; \\
\label{eq:IgE}
&& I_{\rm EQB}=\frac12M_{\rm EQB}[(R_{\rm g})_{\rm EQB}]^2~~; \\
\label{eq:etaT}
&& \eta_{\rm TOV}=\eta_{\rm EQB}=\frac12~~;
\end{lefteqnarray}
where, aiming to preserve continuity, TOV configurations are conceived as
nonrotating disks of the kind considered instead of nonrotating homogeneous
spheres implying
$I_{\rm TOV}=(2/5)M_{\rm TOV}[(R_{\rm g})_{\rm TOV}]^2$.

Parameters related to TOV and EQB configurations (in the above specified
sense), for masses equal to their NS/QS counterparts listed in Table
\ref{t:nqrt}, are presented in Table \ref{t:nqbh}.
\begin{table*}
\caption[par]{Comparison between TOV and EQB configuration parameters (first
and second line of each panel) and additional parameters of EQB configurations
(third line of each panel), for BHs of equal mass with respect to their NS/QS
counterparts.   Parameters are specified on the head of the
Table as: mass, equatorial radius, moment of inertia, profile parameter (first
and second line of each panel), and: angular momentum, specific angular
momentum, spin parameter, period (third line of each panel).   EOSs related to
each panel are (from top to bottom): BCPM, BSk20, BSk21, Shen, CIDDM, CDDM1,
CDDM2, where the first four are applied to NSs and the last three to QSs.
Data are taken or inferred from the parent paper \cite{lia16}.   See text for
further details.}
\label{t:nqbh}
\begin{center}
\begin{tabular}{llll}
\hline
\multicolumn{1}{c}{$M_{\rm TOV}/m_\odot$} &
\multicolumn{1}{c}{$(R_{\rm g})_{\rm TOV}/{\rm km}$} &
\multicolumn{1}{c}{$I_{\rm TOV}/(10^{38}{\rm kg\,m^2})$} &
\multicolumn{1}{c}{$\eta_{\rm TOV}$} \\
\multicolumn{1}{c}{$M_{\rm EQB}/m_\odot$} &
\multicolumn{1}{c}{$(R_{\rm g})_{\rm EQB}/{\rm km}$} &
\multicolumn{1}{c}{$I_{\rm EQB}/(10^{38}{\rm kg\,m^2})$} &
\multicolumn{1}{c}{$\eta_{\rm EQB}$} \\
\multicolumn{1}{c}{$J_{\rm EQB}/(\rm10^{42}kg\,m^2/s)$} &
\multicolumn{1}{c}{$j_{\rm EQB}/({\rm10^{12}m^2/s})$} &
\multicolumn{1}{c}{$a_{\rm EQB}^*$} &
\multicolumn{1}{c}{$T_{\rm EQB}/{\rm ms}$} \\
\hline\noalign{\smallskip}
1.98000D+00 & 5.84635D+00 & 6.72886D$-$01 & 5.00000D$-$01 \\ 
2.33809D+00 & 6.90367D+00 & 1.10797D$+$00 & 5.00000D$-$01 \\ 
4.81136D+00 & 1.03483D+00 & 1.00000D$+$00 & 1.44690D$-$01 \\ 
\hline                                                        
2.17000D+00 & 6.40736D+00 & 8.85778D$-$01 & 5.00000D$-$01 \\ 
2.57642D+00 & 7.60740D+00 & 1.48250D$+$00 & 5.00000D$-$01 \\ 
5.84225D+00 & 1.14032D+00 & 1.00000D$+$00 & 1.59439D$-$01 \\ 
\hline                                                        
2.28000D+00 & 6.73216D+00 & 1.02743D$+$00 & 5.00000D$-$01 \\ 
2.72701D+00 & 8.05204D+00 & 1.75794D$+$00 & 5.00000D$-$01 \\ 
6.54515D+00 & 1.20697D+00 & 1.00000D$+$00 & 1.68758D$-$01 \\ 
\hline                                                        
2.18000D+00 & 6.43689D+00 & 8.98080D$-$01 & 5.00000D$-$01 \\ 
2.59936D+00 & 7.67514D+00 & 1.52246D$+$00 & 5.00000D$-$01 \\ 
5.94676D+00 & 1.15047D+00 & 1.00000D$+$00 & 1.60859D$-$01 \\ 
\hline                                                        
2.09000D+00 & 6.17115D+00 & 7.91379D$-$01 & 5.00000D$-$01 \\ 
2.92295D+00 & 8.63060D+00 & 2.16476D$+$00 & 5.00000D$-$01 \\ 
7.51951D+00 & 1.29369D+00 & 1.00000D$+$00 & 1.80884D$-$01 \\ 
\hline                                                        
2.21000D+00 & 6.52547D+00 & 9.35669D$-$01 & 5.00000D$-$01 \\ 
3.09282D+00 & 9.13216D+00 & 2.56453D$+$00 & 5.00000D$-$01 \\ 
8.41889D+00 & 1.36888D+00 & 1.00000D$+$00 & 1.91396D$-$01 \\ 
\hline                                                        
2.45000D+00 & 7.23412D+00 & 1.27481D$+$00 & 5.00000D$-$01 \\ 
3.44237D+00 & 1.01643D+01 & 3.53605D$+$00 & 5.00000D$-$01 \\ 
1.04295D+01 & 1.52359D+00 & 1.00000D$+$00 & 2.13028D$-$01 \\ 
\noalign{\smallskip}
\hline                             
\end{tabular}                      
\end{center}                       
\end{table*}                       
Masses, gravitational radii, moments of inertia, profile parameters, for TOV
and EQB configurations are listed on the first and second line in each panel.
Angular momenta, specific angular momenta, spin parameters, periods, for EQB
configurations are listed on the third line of each panel.

For maximally rotating $(a_{\rm BH}^*=1)$ BHs, Eq.\,(\ref{eq:JBH}) via
(\ref{eq:Rg}) and (\ref{eq:cons}) reduces to:
\begin{eqnarray*}
&& \frac{J_{\rm BH}}{\rm10^{42}kg\,m^2/s}10^{42}\frac{\rm kg\,m^2}{\rm s}=
\frac{G(M/m_\odot)^2m_\odot^2}c \\
&& \phantom{\frac{J_{\rm BH}}{\rm10^{42}kg\,m^2/s}\rm10^{42}\frac
{\rm kg\,m^2}{\rm s}}=
\frac{C_{\rm G}10^{-11}(M/m_\odot)^2C_\odot^210^
{60}}{C_{\rm c}10^8}\frac{\rm kg\,m^2}{\rm s}~~;
\end{eqnarray*}
which after little calculus reads:
\begin{lefteqnarray}
\label{eq:JBHM}
&& \frac{J_{\rm BH}}{\rm10^{42}kg\,m^2/s}=
\frac{C_{\rm G}C_\odot^2}{10C_{\rm c}}\left(\frac M{m_\odot}\right)^2~~;
\end{lefteqnarray}
where $C_{\rm G}C_\odot^2/(10C_{\rm c})=0.880130$.

Turning Eq.\,(\ref{eq:JBHM}) into logarithms yields:
\begin{lefteqnarray}
\label{eq:lgJM}
&& \log\frac{J_{\rm BH}}{\rm10^{42}kg\,m^2/s}=2\log\frac M{m_\odot}+\log
C_{\rm G}+2\log C_\odot-\log C_{\rm c}-1~~; \\
&& \log C_{\rm G}+2\log C_\odot-\log C_{\rm c}-1=-0.05545~~;
\end{lefteqnarray}
which is a straight line of slope, 2, and intercept, $-$0.05545, on the
logarithmic plane
$\{{\sf O}\log(M/m_\odot)\log[J/(10^{42}{\rm kg\,m^2/s})]\}$.

Let a NS/QS rigidly rotate with assigned frequency, $\Omega=2\pi/T$, for
fixed EOS.   Related maximally rotating configuration is defined by
$(M_{\rm max}, R_{\rm max}, I_{\rm max})$ according to
Eqs.\,(\ref{eq:Mmax})-(\ref{eq:Imax}) \cite{lia16} and, owing to rigid
rotation, angular momentum reads
$J_{\rm max}=I_{\rm max}\Omega=I_{\rm max}2\pi/T$ or, using
Eq.\,(\ref{eq:Imax}):
\begin{lefteqnarray}
\label{eq:Jmax}
&& \frac{J_{\rm max}}{\rm10^{42}kg\,m^2/s}=\frac{2\pi/10}{T/{\rm ms}}\frac
{M_{\rm max}}{m_\odot}\left(\frac{R_{\max}}{\rm km}\right)^2\frac{C_\odot}
{100}\frac a{1+\exp[-kT(1-q/T)]}~~;\quad
\end{lefteqnarray}
where $J_{\rm max}=J_{\rm TOV}=0$ for $\Omega=0$ and $J_{\rm max}=J_{\rm EQB}$
for $\Omega=\Omega_{\rm EQB}$, $T=T_{\rm EQB}$.

\begin{figure*}[t]  
\begin{center}      
\includegraphics[scale=0.8]{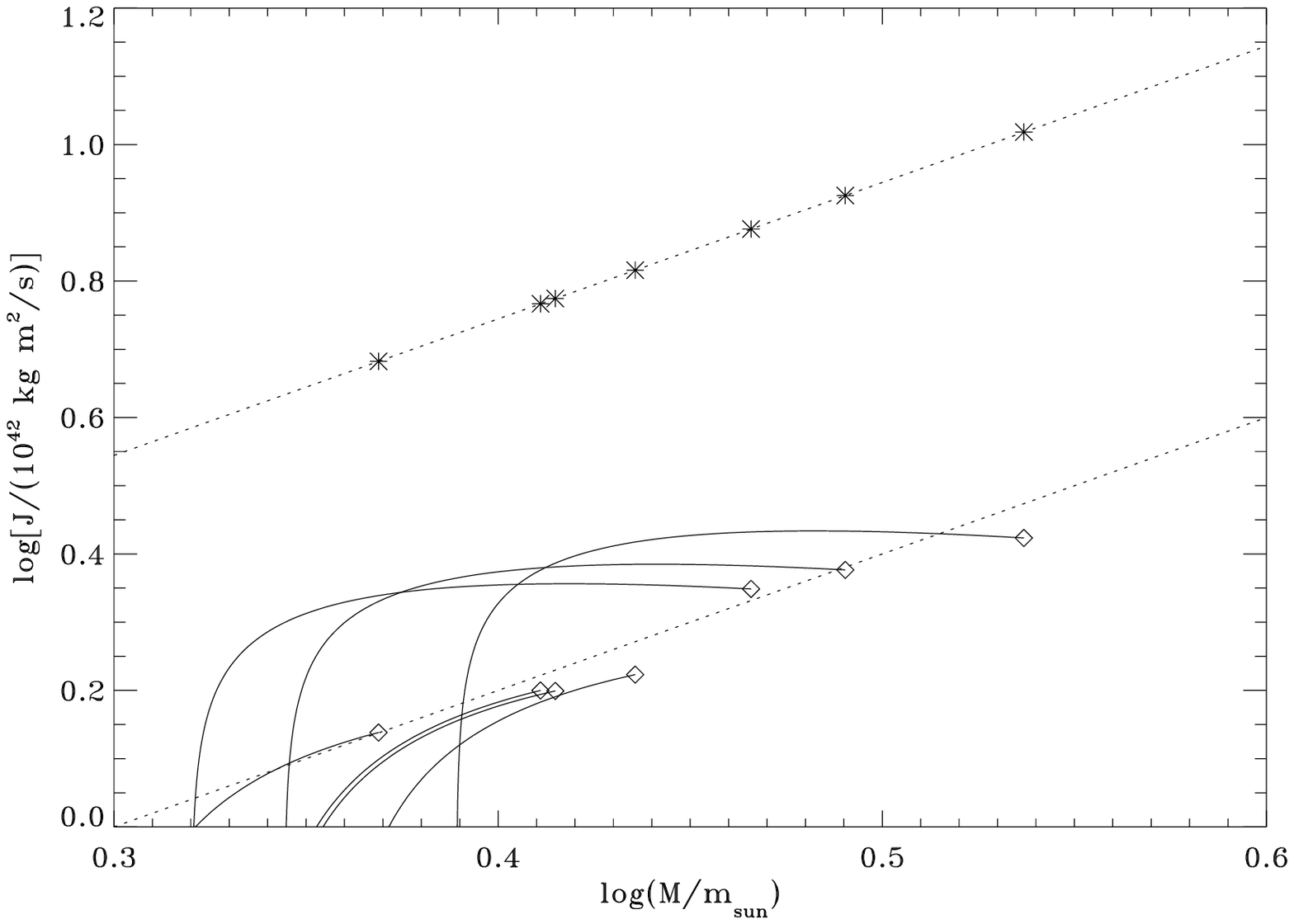}                      
\caption[ddbb]{$J$-$M$ relation in logarithmic plane for EQB configurations
related to NSs/QSs (diamonds) and maximally rotating BHs of equal mass,
$M_{\rm BH}=M_{\rm EQB}$ (asterisks), placed on the dotted straight line of
slope, 2, and intercept, $-$0.05545.   A parallel line (arbitrarily chosen at
$\log a_{\rm BH}^*=-0.5545$ or $a_{\rm BH}^*=0.2789$) is placed downwards to
facilitate comparison with EQB configurations.   Loci of maximum mass
configurations for frequencies up to EQB value are shown as full curves.  With
regard to EQB configurations, EOSs are, from the left to the right: (a) NSs -
BCPM, BSk20, Shen, BSk21; (b) QSs - CIDDM, CDMM1, CDMM2.  See text for further
details.}
\label{f:jvsm}     
\end{center}       
\end{figure*}

Related curves on the logarithmic plane,
$[{\sf O}\log(M/m_\odot)\log(J/10^{42}{\rm kg\,m^2/s})]$, are plotted in
Fig.\,\ref{f:jvsm} where ending points i.e. EQB configurations are shown as
diamonds.
Maximally rotating $(a_{\rm BH}^*=1)$ BHs of equal mass,
$M_{\rm BH}=M_{\rm EQB}$, are shown as asterisks, which are placed on the
dotted straight line, related to Eq.\,(\ref{eq:lgJM}).   A parallel line
(arbitrarily chosen at
$\log a_{\rm BH}^*=-0.5545$ or $a_{\rm BH}^*=0.2789$) is placed downwards to
facilitate comparison with EQB configurations.   Related
EOSs are, from the left to the right with regard to ending points: (a) NSs -
BCPM, BSk20, Shen, BSk21; (b) QSs - CIDDM, CDMM1, CDMM2.

Interestingly, EQB configurations are aligned to a good extent for both NSs
and QSs, but slope is lower with respect to maximally rotating BHs.
Accordingly, EQB configurations can be conceived as maximally rotating
configurations, regardless of EOS.   Related angular momentum is lower with
respect to their maximally rotating BH counterparts by a factor of about
3.5851.
Finally, it is worth of note equatorial velocity of EQB configurations,
$v_{\rm EQB}=\Omega_{\rm EQB}R_{\rm EQB}=2\pi R_{\rm EQB}/T_{\rm EQB}$,
inferred from Table \ref{t:nqrt} and Eq.\,(\ref{eq:Rg}), equals about one
half the light velocity, $v_{\rm EQB}\approx c/2$.

\section{Quark-level fusion in heavy baryons}
\label{quafu}

According to a recent investigation, quark-level fusion with release of energy
can occur as well as nuclear fusion, involving heavy baryons \cite{KaR17}, see
also \cite{Mil17}.   More specifically, (i) fusion between two heavy baryons
$(\Lambda_{\rm c}^+)$ produces a doubly charmed baryon $(\Xi_{\rm cc}^{++})$
and a neutron $(n)$,
$\Lambda_{\rm c}^+\Lambda_{\rm c}^+\to\Xi_{\rm cc}^{++}n$, with
an energy release of 12 MeV; (ii) fusion between a heavy $(\Lambda_{\rm c}^+)$
and a heavier $(\Lambda_{\rm b}^0)$ baryon produces a charmed-bottomed baryon
$(\Xi_{\rm cb}^+)$ and a neutron (n),
$\Lambda_{\rm c}^+\Lambda_{\rm b}^0\to\Xi_{\rm cb}^+n$, with an energy release
of 50 MeV; (iii) fusion between two heavier baryons
$(\Lambda_{\rm b}^0)$ produces a doubly bottomed baryon $(\Xi_{\rm bb}^0)$ and
a neutron $(n)$, $\Lambda_{\rm b}^0\Lambda_{\rm b}^0\to\Xi_{\rm bb}^0n$, with
an energy release of 138 MeV.

In general, a reaction can be written as
$\Lambda_{\rm u}\Lambda_{\rm v}\to\Xi_{\rm uv}N$, where uv = ss, cc, bb, cb,
and $N$ is a nucleon.   The reaction,
$\Lambda_{\rm s}\Lambda_{\rm s}\to\Xi_{\rm ss}N$, is endothermic with an
energy release of $-$23 MeV.   For further details, an
interested reader is addressed to the parent paper \cite{KaR17}.

As discussed therein, energy release is largely determined by binding energy
between heavy quarks.   Related interactions take place via an effective
two-body potential, which implies binding energy is determined by their
reduced mass, $\mu_{\rm red}=m(\Lambda_{\rm u})m(\Lambda_{\rm v})/
[m(\Lambda_{\rm u})+m(\Lambda_{\rm v})]$, where $m(\Lambda_{\rm u})$ and
$m(\Lambda_{\rm v})$ are masses of individual quarks.   It can be seen
energy release, $\Delta E$, linearly depends on reduced mass, $\mu_{\rm red}$,
as \cite{KaR17}:
\begin{lefteqnarray}
\label{eq:DEmr}
&& \frac{\Delta E}{\rm MeV}=-44.95+0.0726\mu_{\rm red}~~;
\end{lefteqnarray}
with regard to the above mentioned quark-level fusion reactions,
$\Lambda_{\rm u}\Lambda_{\rm v}\to\Xi_{\rm uv}N$.

Given a NS/QS, let the following restrictive assumptions hold: (1) a
substantial mass fraction, $M_{\rm uv}/M_{\rm NS/QS}$, is in form of
heavy baryon reagents, $\Lambda_{\rm u}$, u = s, c, b; and (2) if
$\Lambda_{\rm u}\Lambda_{\rm v}\to\Xi_{\rm uv}N$ reactions take place, the
extension to the whole mass, $M_{\rm uv}$, is immediate.   Accordingly, the
number of reactions involving $\Lambda_{\rm u}$, $\Lambda_{\rm v}$, reads:
\begin{lefteqnarray}
\label{eq:NLa}
&& N_{\rm uv}=\frac{M_{\rm uv}}{m(\Lambda_{\rm u})+m(\Lambda_{\rm v})}~~;
\end{lefteqnarray}
where u,v = s, c, b.   The total energy,
$E_{\rm uv}$, released from $\Lambda_{\rm u}\Lambda_{\rm v}\to\Xi_{\rm uv}N$
reactions throughout the whole mass, $M_{\rm uv}$, is:
\begin{lefteqnarray}
\label{eq:Euv}
&& E_{\rm uv}=N_{\rm uv}\Delta E_{\rm uv}~~;
\end{lefteqnarray}
where $\Delta E_{\rm uv}$ is energy released from a single reaction.

Related values are listed in Table \ref{t:LLCN}: mass of heavy baryons,
$m(\Lambda_{\rm u})$, $m(\Xi_{\rm uv})$, involved in reactions,
$\Lambda_{\rm u}\Lambda_{\rm v}\to\Xi_{\rm uv}N$; energy release,
$\Delta E_{\rm uv}$; number of reactions per solar mass of reagents,
$N_{\rm uv}$; energy release per solar mass of reagents, $E_{\rm uv}$; for the
following combinations of heavy quarks: uv = ss, cc, bb, cb.
\begin{table*}
\caption[par]{Mass of heavy baryons, $m(\Lambda_{\rm u})$, $m(\Xi_{\rm uv})$,
involved in reactions, $\Lambda_{\rm u}\Lambda_{\rm v}\to\Xi_{\rm uv}N$, where
$N$ is a nucleon; energy release, $\Delta E_{\rm uv}$; number of reactions
per solar mass of reagents,
$N_{\rm uv}=m_\odot/[m(\Lambda_{\rm u})+m(\Lambda_{\rm v})]$; energy release
per solar mass of reagents, $E_{\rm uv}=N_{\rm uv}\Delta E_{\rm uv}$; for the
following combinations of heavy quarks: uv = ss, cc, bb, cb.   In the last
case, reagent masses can be read on related left columns.   Values on the
first three rows are taken from the parent paper \cite{KaR17} and converted
from MeV to kg (mass) and J (energy), 1 MeV = $1.78266270\cdot10^{-30}$kg =
$1.60217733\cdot10^{-13}$J.   See text for further details.}
\label{t:LLCN}
\begin{center}
\begin{tabular}{lllll}
\hline
\multicolumn{1}{c}{reaction:} &
\multicolumn{1}{c}{uv = ss} &
\multicolumn{1}{c}{uv = cc} &
\multicolumn{1}{c}{uv = bb} &
\multicolumn{1}{c}{uv = cb} \\
\multicolumn{1}{c}{parameter} &
\multicolumn{1}{c}{} &
\multicolumn{1}{c}{} &
\multicolumn{1}{c}{} &
\multicolumn{1}{c}{} \\
\hline\noalign{\smallskip}
$m(\Lambda_{\rm u})$/kg  & $+$1.9989E$-$27 & 4.0761E$-$27 & 1.0018E$-$26 & (left values)  \\
$m(\Xi_{\rm uv})$/kg     & $+$2.3440E$-$27 & 6.4557E$-$27 & 1.8115E$-$26 & 1.2331E$-$26 \\
$\Delta E_{\rm uv}$/J    & $-$3.7010E$-$12 & 1.9226E$-$12 & 2.2110E$-$11 & 8.0109E$-$12 \\
$N_{\rm uv}$             & $+$4.9991E$+$56 & 2.4393E$+$56 & 9.9249E$+$55 & 1.4109E$+$56 \\
$E_{\rm uv}$/J           & $-$1.8502E$+$45 & 4.6898E$+$44 & 2.1784E$+$45 & 1.1307E$+$45 \\
\noalign{\smallskip}
\hline                             
\end{tabular}                      
\end{center}                       
\end{table*}                       

With regard to a single reaction, mass and energy values are taken from the
parent paper \cite{KaR17} and converted from MeV to kg (mass) and J (energy),
1 MeV = $1.78266270\cdot10^{-30}$kg = $1.60217733\cdot10^{-13}$J.
To optimize $\Delta E_{\rm ss}$, $\Xi=\Xi^0(\rm ssu)$ and $N=n$ are
considered, as $m[\Xi^-(\rm ssd)]$ is 7 MeV larger than $m[\Xi^0(\rm ssu)]$
\cite{KaR17}.   For
further details, an interested reader is addressed to the parent paper
\cite{KaR17}, where uncertainties in $m(\Xi)$ and $\Delta E_{\rm uv}$ are also
mentioned.

An inspection of Table \ref{t:LLCN} discloses total energy released from a
solar mass of reagents ranges as
$4.7\cdot10^{44}\appleq E_{\rm uv}/{\rm J}\appleq2.2\cdot10^{45}$ according to
quark families present in reagents.  Similarly, the total energy absorbed by a
solar mass of reagents in endothermic reactions amounts to
$E_{\rm ss}/{\rm J}\approx-1.8\cdot10^{45}$.

On the other hand, kinetic
energy released from typical supernova explosions range as
$10^{43}\appleq E_{\rm SN}/{\rm J}\appleq10^{45}$, 1\,J = $10^7$\,erg, e.g.,
\cite{mal07} \cite{Kus18}, and total energy released from core-collapse
supernovae can grow up to a few
$10^{46}$ mainly due to neutrino emission e.g., \cite{LaY89} \cite{LaP16}
\cite{GVV17}.   Accordingly, energy carried by neutrinos can safely be
considered as equal to binding energy, $E_{\rm B}$, released in core collapse
to form NS/QS e.g., \cite{LaY89} \cite{LaP07}.

In general, the binding energy of a discrete matter distribution is defined as
the difference in mass between matter distribution at infinite dispersion (the
particle mass, $N\overline m$) and the gravitational mass (the system mass,
$M$) e.g., \cite{LaP16}.   In Newtonian gravity, the binding energy of a
homogeneous sphere is the opposite of the potential energy e.g., \cite{LaP07}
or in explicit form:
\begin{lefteqnarray}
\label{eq:EBC}
&& E_{\rm B}=\zeta_{\rm pot}\frac{GM^2}R=\zeta_{\rm pot}\frac{GM^2}{R_{\rm g}}
\frac{R_{\rm g}}R=\zeta_{\rm pot}\frac{GM^2}{2GMc^{-2}}\frac{R_{\rm g}}R=
\zeta_{\rm pot}\beta^*Mc^2~~;
\end{lefteqnarray}
where $\zeta_{\rm pot}$ is an increasing function of central to mean density
ratio, $\rho_{\rm c}/\overline\rho$, and $\zeta_{\rm pot}=3/5$ for a
homogeneous sphere, $\overline\rho=\rho_{\rm c}$.   The special case of
spherical-symmetric polytropes reads e.g., \cite{Cha39} Chap.\,IV \S7:
\begin{lefteqnarray}
\label{eq:EBP}
&& E_{\rm B}=\frac3{5-n}\beta^*Mc^2~~;
\end{lefteqnarray}
ranging from homogeneous $(n=0)$ to Roche or infinitely extended $(n=5)$
configurations.

With regard to NS/QS remnants, an empirical relation inferred from EOSs
consistent with an upper mass limit, $M_{\rm max}\appgeq1.65m_\odot$, is
\cite{LaP01} \cite{LaP07} \cite{LaP16}:
\begin{leftsubeqnarray}
\slabel{eq:EBNa}
&& \frac{E_{\rm B}}{Mc^2}=\frac35\beta^*\left(1-\frac12\beta^*\right)^{-1}~~;
\\
\slabel{eq:EBNb}
&& \Delta\frac{E_{\rm B}}{Mc^2}=\frac1{20}\beta^*\left(1-\frac12\beta^*\right)
^{-1}~~;
\label{seq:EBN}
\end{leftsubeqnarray}
where $0.10\appleq\beta^*\le\beta_{\rm max}^*$, $\beta_{\rm max}^*\approx0.35$
e.g., \cite{LaP07}.

\begin{figure*}[t]  
\begin{center}      
\includegraphics[scale=0.8]{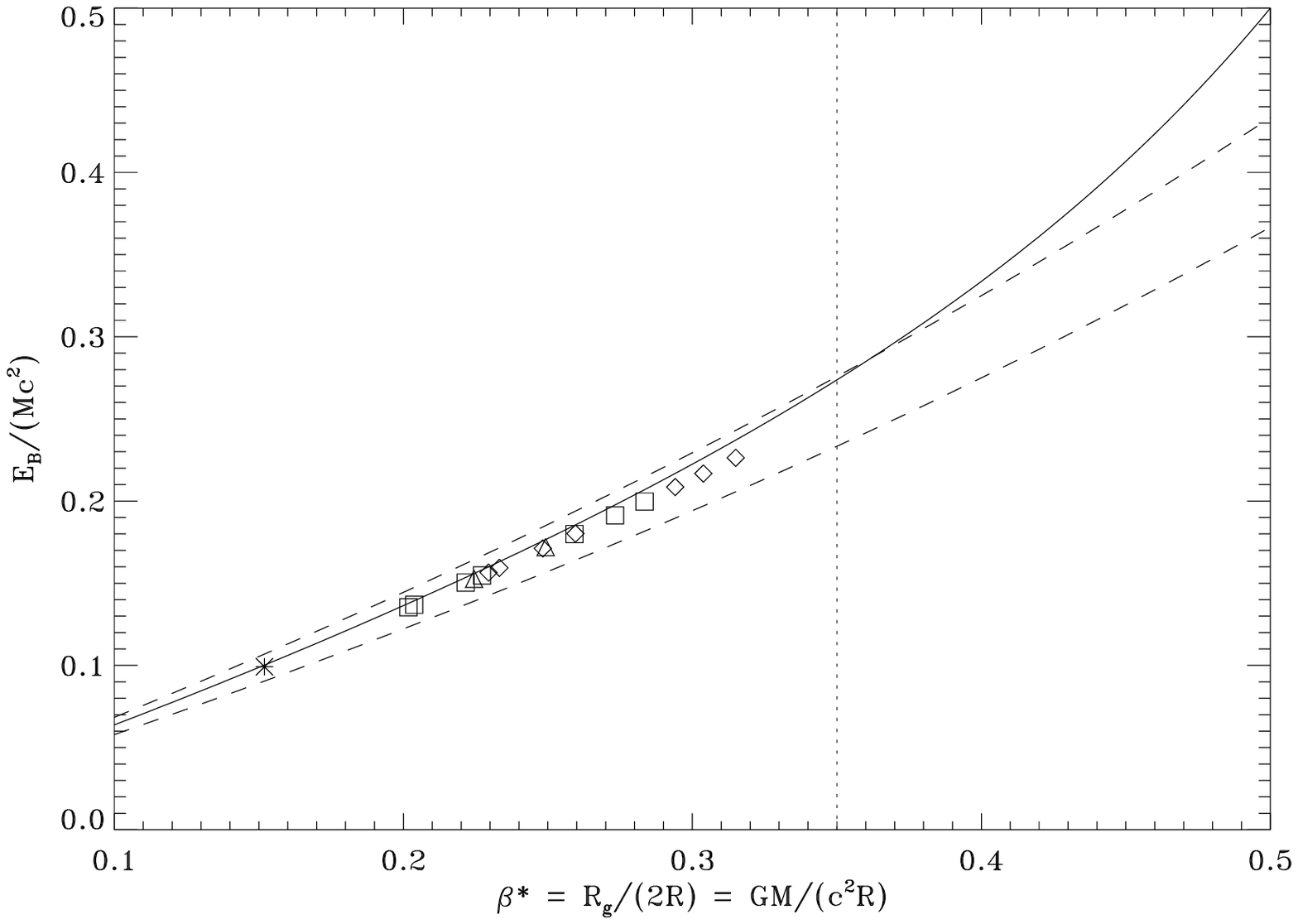}                      
\caption[ddbb]{Binding energy of TOV (diamonds) and EQB (squares)
configurations, expressed via an empirical relation,
$E_{\rm B}/(Mc^2)=(0.6\mp0.05)\beta^*(1-0.5\beta^*)^{-1}$ \cite{LaP01}, where 
the tolerance is marked by the dashed curves.   Also plotted for comparison
are PSR J0737-3039A (asterisk) \cite{ROq16}, and PSR J0348+0432 (triangles)
related to lower (left) and upper (right) mass limit \cite{Zha17}.   The full
curve results from the addition of an extra term,
$(E_{\rm B}+E_{\rm B}^\prime)/(Mc^2)=\{0.6\beta^*+0.075\exp[-8(0.5-\beta^*)]
\}(1-0.5\beta^*)^{-1}$, which satisfies the conditon,
$(E_{\rm B}+E_{\rm B}^\prime)/(Mc^2)=0.5$, in the relativistic limit,
$\beta^*=0.5$.   The maximum compactness parameter for NS/QS, taken as
$\beta_{\rm max}^*=0.35$ e.g., \cite{LaP07}, is marked as a vertical dotted
line.   See text for further details.}
\label{f:bele}     
\end{center}       
\end{figure*}

With regard to above considered NS/QS remnants, compactness parameter,
$\beta^*$, can be inferred from Tables \ref{t:nqrt} and \ref{t:nqbh}, where
$R$ and $R_{\rm g}$
are listed, respectively.   Related binding energy, $E_{\rm B}/(Mc^2)$, can be
estimated via Eq.\,(\ref{seq:EBN}).   Results are plotted in
Fig.\,\ref{f:bele} for TOV (diamonds) and EQB (squares) configurations.
The tolerance, $\mp\Delta[E_{\rm B}/(Mc^2)]$, is represented by
the dashed band.   Also plotted for
comparison are PSR J0737-3039A (asterisk) \cite{ROq16} and PSR J0348+0432
(triangles) related to lower (left) and upper (right) mass limit \cite{Zha17}.

Under the restrictive assumption NS/QS binding energy for fixed EOS depends
only on compactness parameter, as:
\begin{lefteqnarray}
\label{eq:EBB}
&& E_{\rm B}=\beta^*f(\beta^*)Mc^2~~;
\end{lefteqnarray}
the classical counterpart, expressed by Eq.\,(\ref{eq:EBC}), is reproduced
provided $\zeta_{\rm pot}=f(\beta^*)$.   The special case of polytropes reads
$f(\beta^*)=3/(5-n)$.   On the other hand, $f(\beta^*)=3/[5(1-\beta^*/2)]$ via
Eq.\,(\ref{eq:EBNa}).   Accordingly, compactness parameter may be related to 
polytropic index as:
\begin{lefteqnarray*}
&& \frac3{5-n}=\frac35\left(1-\frac12\beta^*\right)^{-1}~~;\quad
\frac{5-n}5=1-\frac12\beta^*~~;\quad\frac n5=\frac12\beta^*~~;
\end{lefteqnarray*}
which is equivalent to:
\begin{lefteqnarray}
\label{eq:betn}
&& n=\frac52\beta^*~~;
\end{lefteqnarray}
and keeping in mind $0.10\appleq\beta^*\appleq0.35$ for NS/QS e.g.,
\cite{LaP07}, related polytropic indexes read $1/4\appleq n\appleq7/8$.
Classical $(\beta^*=0)$ and relativistic $(\beta^*=1/2)$ limit imply $n=0$ and
$n=5/4$, respectively.

Though stable NS/QS configurations cannot exist for
$\beta^*>\beta_{\rm max}^*\approx0.35$
e.g., \cite{LaP07}, binding energy extrapolation to relativistic limit could
be of some interest.   Using the empirical fit, $f(1/2)=(3/5)/(3/4)=4/5$
and $E_{\rm B}(1/2)=(2/5)Mc^2$ via Eqs.\,(\ref{seq:EBN}) and (\ref{eq:EBB}).
An improved fit, based on an array of piecewise polytrope EOSs consistent with
an upper mass limit, $M_{\rm max}\appgeq1.97m_\odot$, and limited to
$\beta\appgeq0.1$ \cite{SLB16}, yields
$E_{\rm B}(1/2)=(1/2)(0.564+0.521/2)Mc^2=0.41225Mc^2$.    The value,
$E_{\rm B}(1/2)=(1/2)Mc^2$, would imply binding energy equals one half rest
energy at BH limit and larger values erase baryons as distinct particles.

An empirical fit can be arranged to satisfy the above mentioned boundary
condition with the addition of a new term, $E_{\rm B}^\prime(\beta^*)$, which
is consistent with the tolerance of the original fit for 
$\beta^*<\beta_{\rm max}^*\approx0.35$.   Accordingly, the following relations
hold:
\begin{lefteqnarray}
\label{eq:EpB1}
&& \frac{E_{\rm B}^\prime(\beta_{\rm max}^*)}{Mc^2}\le
\Delta\frac{E_{\rm B}(\beta_{\rm max}^*)}{Mc^2}~~; \\
\label{eq:EcB2}
&& \frac{E_{\rm B}(0)+E_{\rm B}^\prime(0)}{Mc^2}=0~~; \\
\label{eq:EpB2}
&& \frac{E_{\rm B}(1/2)+E_{\rm B}^\prime(1/2)}{Mc^2}=\frac12~~;
\end{lefteqnarray}
where, for reasons of simplicity, attention shall be restricted to the fit
expressed by Eq.\,(\ref{seq:EBN}) and to an exponential function with negative 
argument for $E_{\rm B}$ and $E_{\rm B}^\prime$, respectively.

Under the above mentioned restrictions, the arranged fit takes the expression:
\begin{lefteqnarray}
\label{eq:EpB3}
&& \frac{E_{\rm B}(\beta^*)+E_{\rm B}^\prime(\beta^*)}{Mc^2}=\left\{\frac35
\beta^*+\frac3{20}\beta^*\exp\left[-\kappa\left(\frac12-\beta^*\right)
\right]\right\}\left(1-\frac12\beta^*\right)^{-1};\qquad
\end{lefteqnarray}
accordingly, Eqs.\,(\ref{eq:EpB1}) and (\ref{eq:EpB2}) read:
\begin{lefteqnarray}
\label{eq:EpB4}
&& \frac3{20}\beta_{\rm max}^*\exp\left[-\kappa\left(\frac12-\beta_{\rm max}
^*\right)\right]\left(1-\frac12\beta_{\rm max}^*\right)^{-1}\le\frac1{20}
\beta_{\rm max}^* \nonumber \\
&& \phantom{\frac3{20}\beta_{\rm max}^*\exp\left[-\kappa\left(\frac12-\beta_
{\rm max}^*\right)\right]\left(1-\frac12\beta_{\rm max}^*\right)^{-1}\le}
\times
\left(1-\frac12\beta_{\rm max}^*\right)^{-1}; \\
\label{eq:EpB5}
&& \frac{E_{\rm B}(1/2)+E_{\rm B}^\prime(1/2)}{Mc^2}=\left[\frac35\frac12
+\frac3{20}\frac12\exp(0)\right]\left(1-\frac12\frac12\right)^{-1}=\frac{15}
{40}\frac43=\frac12~~;\qquad
\end{lefteqnarray}
where, setting $\beta_{\rm max}^*=0.35$, the inequality, Eq.\,(\ref{eq:EpB4}),
reduces to:
\begin{lefteqnarray*}
&& \exp(-0.15\kappa)\le\frac{20}3\frac1{20}=\frac13~;~~
-0.15\kappa\le-\ln3~;~~\kappa\ge\frac{20}3\ln3\approx7.324082~;\qquad
\end{lefteqnarray*}
where a suitable choice reads $\kappa=8$.

Accordingly, the arranged fit via Eq.\,(\ref{eq:EpB3}) takes the explicit
expression:
\begin{lefteqnarray}
\label{eq:EpB6}
&& \frac{E_{\rm B}(\beta^*)+E_{\rm B}^\prime(\beta^*)}{Mc^2}=\frac35\beta^*
\left\{1+\frac14\exp\left[-8\left(\frac12-\beta^*\right)\right]\right\}
\left(1-\frac12\beta^*\right)^{-1}~~;\qquad
\end{lefteqnarray}
which is plotted in Fig.\,\ref{f:bele} as a full curve.

NS/QS binding energy amounts to a few $10^{46}$\,J e.g., \cite{LaY89}
\cite{LaP16} \cite{GVV17} \cite{BoL18}, which exceeds energy released via
quark-level fusion by a factor of about 10.   In other words, NS/QS binding
energy changes of about 10\% at most during quark-level fusion, preventing
either supernova explosion or collapse into submassive BH.

\section{Discussion}
\label{disc}

The transition from NS to BH remnants shows a desert within the mass range,
$2\appleq M/m_\odot\appleq4$ e.g., \cite{faa11} as illustrated in
Fig.\,\ref{f:lico}.   Leaving aside occasional fallback from supernovae,
massive $(M\appgeq8m_\odot)$ ZAMS progenitors seem to be excluded for the
following reasons: (i) the lower mass limit of BH progenitors equals
$15m_\odot$ and, (ii) the lower mass limit of BHs from massive progenitors
equals $4m_\odot$; as shown for solar metallicities in a recent investigation
\cite{RSO18}.   In this view, different processes must be considered for the
formation of subHe-core (in particular, submassive)
BHs.

Compact remnants in close binaries, such as WD-WD, WD-NS, NS-NS, merge after
orbital decay via gravitational radiation, as in the recent event GW 170817
where the inferred mass is $2.73_{-0.01}^{+0.04}m_\odot$ \cite{aba17}.
According to
initial masses, the merger product could be WD, NS, subHe-core BH, or
supernova explosion leaving no remnant.   Binary systems hosting NS and donor
MS star could yield submassive BH via mass accretion onto NS and subsequent
instability, where BH presence should be inferred as mass accretion goes on.
But submassive BHs in binary systems have not been detected up today.

Ultramassive, rapidly rotating or highly magnetized NSs, where spin rate or
magnetic field are progressively reduced, finally collapse into submassive
BHs e.g., \cite{lia16} \cite{haa16} \cite{RMW18}.   Energy release from
quark-level reactions is found to be less than NS binding energy,
$E_{\rm B}\approx1$-$3\cdot10^{46}\,$J, by a factor of about 10 as shown in
Table \ref{t:LLCN}, which rules out the possibility of supernova explosion
from slowly rotating or lowly magnetized, ultramassive NSs.

The rarity of the above mentioned events, i.e. orbital decay in close WD-WD,
WD-NS, NS-NS binaries; spin decrease or magnetic field attenuation in
ultramassive NSs; fallback from supernovae yielding subHe-core BHs; by itself
could provide an explanation to the desert between NS and BH remnants.

On the other hand, biases and/or selection effects cannot still be excluded,
keeping in mind isolated BHs are difficult to be detected.   Concerning X-ray
transients with low-mass donors, it has been argued that understimates of the
inclination, by $10^\circ$ at least, could significantly reduce BH mass
estimates, filling the gap between the low end of BH mass distribution and the
maximum theoretical NS mass \cite{kra12}.

The rarity of mildly massive $(4\appleq M/m_\odot\appleq6)$ BH remnants is
somewhat surprising.   If related ZAMS progenitors exhibit $M\appgeq15m_\odot$
\cite{RSO18} and the initial mass function is decreasing with increasing mass,
BH remnants would privilege the lower end of mass distribution, contrary to
current data, as shown in Fig.\,\ref{f:lico}.   The above considerations hold
for isolated ZAMS progenitors, while massive stars are mainly born as close
binaries, which could inhibit the formation of mildly massive BH remnants.

Though heavy baryons can undergo quark-level fusion \cite{KaR17}, energy
released is lower than NS binding energy by a factor of about 10, which
prevents supernova explosions as in supramassive WDs.   Then rapidly
spinning or highly magnetized, ultramassive NSs would inescapably collapse
into submassive BHs via spin or magnetic field reduction, but no clear
evidence for these events has been found up today.

An upper BH mass limit of about $16m_\odot$, inferred from stellar evolution
\cite{RSO18}, would imply BH-BH merger products via orbital decay in close
binaries, of about $32m_\odot$ at most for ZAMS progenitors of solar
metallicity.   On the other hand, merger products
up to about $60m_\odot$ have been inferred from gravitational radiation e.g.,
\cite{aba16}, contrary to the above prediction unless related ZAMS progenitors
exhibit etremely low metallicity, $Z\approx10^{-4}Z_\odot$ \cite{SuW14}, or
strong magnetic fields \cite{pea17}, as pointed out in a recent investigation
\cite{RSO18}.

\section{Conclusion}
\label{conc}

The occurrence of a desert between NS and BH remnants was reviewed using
a set of well-determined masses form different sources, as listed in Tables
\ref{t:mabda}-\ref{t:mabh} and plotted in Fig.\,\ref{f:lico}.   Leaving aside
biases and selection effects, special effort was devoted to two specific
points, namely (i) comparison between physical parameters related to NS/QS
(TOV and EQB) and BH (nonrotating and maximally rotating)
configurations of equal mass, respectively; and, (ii) occurrence of supernova
explosions via quark-level reactions in heavy baryons, leaving no remnant.

Concerning point (i) above, NS/QS physical parameters were taken or inferred
from a recent investigation \cite{lia16} for 4 (NS) and 3 (QS) physically
motivated EOSs, as listed in Table \ref{t:nqrt}.   Related BH physical
parameters were determined, as listed in Table \ref{t:nqbh}.

With regard to $\eta$-$\beta^*$
relation, TOV configurations were found to be consistent or marginally
inconsistent with empirical fits inferred from sets of reliable EOSs
\cite{LaP01} \cite{GVV17}; conversely, EQB configurations were found to be
largely inconsistent (but displaced on a parallel sequence) for NSs and
consistent for QSs, as shown in Fig.\,\ref{f:beta}.

With regard to $J$-$M$ relation, EQB configurations were found to be located
near a straight line of similar slope in comparison to maximally rotating
BHs, but shifted downwards due to lower angular momentum by a factor of about
3.5851, as shown in Fig.\,\ref{f:jvsm}.

Concerning point (ii) above, energy released from quark-level reactions
involving heavy baryons (per solar mass of reagents) was determined using
results by a recent investigation \cite{KaR17}, as listed in Table
\ref{t:LLCN}.
Even if NSs are entirely made of heavy baryons, total energies released via
above mentioned reactions were found to be still lower by a factor of
about 10 with respect to binding energy, which inhibits NS supernova
explosions as in supramassive WDs.

A definitive answer to the existence of a desert between NS and
BH remnants is related to the following questions.
\begin{description}
\item[(1)~] Could ultramassive NSs actually collapse into submassive BHs?
\item[(2)~] Could accreting NSs from MS donors in close orbits actually
collapse into submassive BHs?
\item[(3)~] Are there biases and/or selection effects which prevent to infer
the presence of submassive BHs?
\item[(4)~] Are there biases and/or selection effects which prevent to infer
the presence of mildly massive BHs?
\end{description}
To this respect, increasingly refined theoretical and observational techniques
are needed.

\section*{Acknowledgements}
This research has made use of the Exoplanet Orbit Database and the Exoplanet
Data Explorer at exoplanets.org.

\appendix
\section*{Appendix}

\section{Masses of stellar remnants}
\label{a:remma}

Stellar remnants with well determined masses i.e. percent error
$\Delta^\%m=100(\Delta M/M)<10$ or $\Delta M/M<0.1$, shown in
Fig.\,\ref{f:lico}, are listed below in
Tables \ref{t:mabda}-\ref{t:mabdb}, \ref{t:mams}, \ref{t:mawda}-\ref{t:mawdc},
\ref{t:mans}, \ref{t:mabh}, for BDs, MS dwarfs, WDs, NSs, BHs, respectively.
Owing to the paucity of known BH remnants at present, data affected by larger
percent error are also listed in Table \ref{t:mabh}.
\begin{table*}
\caption[par]{High-precision ($\Delta M/M<0.1$) masses and uncertainties for
$N=41$
brown dwarf (BD) stars hosted in binary systems.  Masses are in solar units.}
\label{t:mabda}
\begin{center}
\begin{tabular}{lllll}
\hline
\multicolumn{1}{c}{system} &
\multicolumn{1}{c}{$M\quad$} &
\multicolumn{1}{c}{$+\Delta M\quad$} &
\multicolumn{1}{c}{$-\Delta M\quad$} &
\multicolumn{1}{c}{ref} \\
\hline\noalign{\smallskip}
CoRoT-3 b                & 0.02081 & 0.00095  & 0.00095  & \cite{exo18} \\
CoRoT-15                 & 0.0604  & 0.0039   & 0.0039   & \cite{Sch18} \\
CoRoT-27 b               & 0.00990 & 0.00073  & 0.00073  & \cite{exo18} \\
CoRoT-33                 & 0.056   & 0.0016   & 0.0017   & \cite{Sch18} \\
gamma Leo A b            & 0.00990 & 0.00079  & 0.00079  & \cite{exo18} \\
EPIC 201702477           & 0.0639  & 0.0016   & 0.0016   & \cite{Sch18} \\
EPIC 219388192           & 0.0348  & 0.00009  & 0.00009  & \cite{Sch18} \\
epsilon Indi B           & 0.0716  & 0.0008   & 0.0008   & \cite{dia18} \\
epsilon Indi C           & 0.0669  & 0.00064  & 0.00064  & \cite{dia18} \\
LUH 16 A                 & 0.03199 & 0.00030  & 0.00028  & \cite{LaS18} \\
LUH 16 B                 & 0.02725 & 0.00025  & 0.00024  & \cite{LaS18} \\
HAT-P-13 c               & 0.01362 & 0.00066  & 0.00066  & \cite{exo18} \\
HD 4747 B                & 0.0668  & 0.0015   & 0.0015   & \cite{pea18} \\
HD 217786 b              & 0.0126  & 0.00109  & 0.00109  & \cite{exo18} \\
HD 168443 c              & 0.01660 & 0.00055  & 0.00055  & \cite{exo18} \\
HD 97233 b               & 0.0189  & 0.00142  & 0.00142  & \cite{exo18} \\
HD 162020 b              & 0.01452 & 0.00049  & 0.00049  & \cite{exo18} \\
HD 22781 b               & 0.01321 & 0.00047  & 0.00047  & \cite{exo18} \\
HD 180314 b              & 0.0216  & 0.00168  & 0.00168  & \cite{exo18} \\
HD 202206 b              & 0.01606 & 0.00065  & 0.00065  & \cite{exo18} \\
HD 16760 b               & 0.01269 & 0.000153 & 0.000153 & \cite{exo18} \\
HD 131664 b              & 0.0175  & 0.00124  & 0.00124  & \cite{exo18} \\
HD 38529 c               & 0.01170 & 0.0004   & 0.0004   & \cite{exo18} \\
HD 136118 b              & 0.1115  & 0.0004   & 0.0004   & \cite{exo18} \\
HD 38801 b               & 0.00956 & 0.00041  & 0.00041  & \cite{exo18} \\
HD 39091 b               & 0.00963 & 0.00036  & 0.00036  & \cite{exo18} \\
HD 106270 b              & 0.01058 & 0.00083  & 0.00083  & \cite{exo18} \\
HD 156846 b              & 0.01051 & 0.00036  & 0.00036  & \cite{exo18} \\
HD 114762 b              & 0.01111 & 0.00074  & 0.00074  & \cite{exo18} \\
HD 10069                 & 0.00970 & 0.00035  & 0.00035  & \cite{Sch18} \\
KELT-1                   & 0.02613 & 0.00088  & 0.00088  & \cite{Sch18} \\
Kepler-39 b              & 0.01735 & 0.00069  & 0.00069  & \cite{exo18} \\
KOI-205                  & 0.0389  & 0.0014   & 0.0011   & \cite{Sch18} \\
KOI-415                  & 0.05932 & 0.00257  & 0.00257  & \cite{Sch18} \\
TYC 3727-1064-1          & 0.01247 & 0.00062  & 0.00062  & \cite{Sch18} \\
WASP-18 b                & 0.00961 & 0.00032  & 0.00032  & \cite{exo18} \\
WASP-30                  & 0.0597  & 0.0012   & 0.0012   & \cite{Sch18} \\
\noalign{\smallskip}
\hline                             
\end{tabular}                      
\end{center}                       
\end{table*}                       
\begin{table*}
\caption[par]{High-precision ($\Delta M/M<0.1$) masses and uncertainties for
$N=41$ brown dwarf (BD) stars hosted in binary systems.  Masses are in solar
units.   Continuation of Table \ref{t:mabda}.}
\label{t:mabdb}
\begin{center}
\begin{tabular}{lllll}
\hline
\multicolumn{1}{c}{system} &
\multicolumn{1}{c}{$M\quad$} &
\multicolumn{1}{c}{$+\Delta M\quad$} &
\multicolumn{1}{c}{$-\Delta M\quad$} &
\multicolumn{1}{c}{ref} \\
\hline\noalign{\smallskip}
WASP-128 b               & 0.0358  & 0.0008   & 0.0008   & \cite{hoa18} \\
XO-3 b                   & 0.01246 & 0.00042  & 0.00042  & \cite{exo18} \\
10 Del b                 & 0.00983 & 0.00034  & 0.00034  & \cite{exo18} \\
11 Com b                 & 0.0154  & 0.00147  & 0.00147  & \cite{exo18} \\
\noalign{\smallskip}
\hline                             
\end{tabular}                      
\end{center}                       
\end{table*}                       
\begin{table*}
\caption[par]{High-precision ($\Delta M/M<0.1$) masses and uncertainties for
$N=19$
main sequence (MS) dwarf stars hosted in binary systems.  Masses are in solar
units.}
\label{t:mams}
\begin{center}
\begin{tabular}{lllll}
\hline
\multicolumn{1}{c}{system} &
\multicolumn{1}{c}{$M\quad$} &
\multicolumn{1}{c}{$+\Delta M\quad$} &
\multicolumn{1}{c}{$-\Delta M\quad$} &
\multicolumn{1}{c}{ref} \\
\hline\noalign{\smallskip}
EBLM J0555-57A            & 0.0813  & 0.0037  & 0.0038  & \cite{Sch18} \\
HATS550-016               & 0.1098  & 0.0057  & 0.0048  & \cite{Sch18} \\
HAT-TR-205-013            & 0.124   & 0.0095  & 0.0095  & \cite{Sch18} \\
Kepler-503 b              & 0.075   & 0.003   & 0.003   & \cite{cab18} \\
KIC 1571511               & 0.143   & 0.0038  & 0.0048  & \cite{Sch18} \\
KOI-189                   & 0.074   & 0.0032  & 0.0032  & \cite{Sch18} \\
KOI-686                   & 0.0987  & 0.0046  & 0.0046  & \cite{Sch18} \\
OGLE-TR-122               & 0.092   & 0.0086  & 0.0086  & \cite{Sch18} \\
SDSS J010448.46+153501.8  & 0.0860  & 0.0015  & 0.0015  & \cite{zab17} \\
SDSS J125637.13-022452.4  & 0.0833  & 0.0015  & 0.0015  & \cite{zaa17} \\
ULAS J020858.62+020657.0  & 0.0827  & 0.0015  & 0.0015  & \cite{zha18} \\
ULAS J135058.86+081506.8  & 0.0833  & 0.0015  & 0.0015  & \cite{zha18} \\
ULAS J151913.03-000030.0  & 0.0811  & 0.0015  & 0.0015  & \cite{zaa17} \\
ULAS J230711.01+014447.1  & 0.0822  & 0.0015  & 0.0015  & \cite{zha18} \\
TYC 7760-484-1            & 0.0911  & 0.0024  & 0.0018  & \cite{Sch18} \\
1SWASPJ234318.41+295556.5 & 0.0955  & 0.0067  & 0.0067  & \cite{Sch18} \\
2MASS J05325346+8246465   & 0.0802  & 0.0015  & 0.0015  & \cite{zaa17} \\
2MASS J06164006-6407194   & 0.0805  & 0.0015  & 0.0015  & \cite{zaa17} \\
2MASS J16262034+3925190   & 0.0828  & 0.0015  & 0.0015  & \cite{zaa17} \\
\noalign{\smallskip}                           
\hline                                         
\end{tabular}                                  
\end{center}
\end{table*}
\begin{table*}
\caption[par]{High-precision ($\Delta M/M<0.1$) masses and uncertainties for
$N=51$ white dwarf (WD) stars hosted in binary systems and $N=25$ nearby
stars belonging to the 25-pc WD sample \cite{sua17}.  Masses are in solar
units.}
\label{t:mawda}
\begin{center}
\begin{tabular}{lllll}
\hline
\multicolumn{1}{c}{system} &
\multicolumn{1}{c}{$M\quad$} &
\multicolumn{1}{c}{$+\Delta M\quad$} &
\multicolumn{1}{c}{$-\Delta M\quad$} &
\multicolumn{1}{c}{ref} \\
\hline\noalign{\smallskip}
CSS 080502       & 0.4756  & 0.0036  & 0.0036  & \cite{paa17} \\  
CSS 09704        & 0.4164  & 0.0356  & 0.0356  & \cite{paa17} \\
CSS 21357        & 0.6579  & 0.0097  & 0.0097  & \cite{paa17} \\
CSS 40190        & 0.4817  & 0.0077  & 0.0077  & \cite{paa17} \\
CSS 41177 A      & 0.3780  & 0.0230  & 0.0230  & \cite{boa14} \\
CSS 41177 B      & 0.3160  & 0.0110  & 0.0110  & \cite{boa14} \\
GD 50            & 1.28    & 0.02    & 0.02    & \cite{gaa18} \\
GK Vir           & 0.5618  & 0.0142  & 0.0142  & \cite{paa12} \\
KIC 08626021     & 0.570   & 0.005   & 0.005   & \cite{gia18} \\
NN Ser           & 0.5354  & 0.0117  & 0.0117  & \cite{paa10} \\
PG1258+593       & 0.54    & 0.01    & 0.01    & \cite{gia10} \\
Procyon B        & 0.592   & 0.006   & 0.006   & \cite{boa15} \\
PSR B1855+09     & 0.238   & 0.013   & 0.012   & \cite{foa16} \\
PSR B2303+46     & 1.3     & 0.1     & 0.1     & \cite{vKK99} \\
PSR J0337+1715 A & 0.19751 & 0.00015 & 0.00015 & \cite{raa14} \\
PSR J0337+1715 B & 0.4101  & 0.0003  & 0.0003  & \cite{raa14} \\
PSR J0348+0432   & 0.172   & 0.003   & 0.003   & \cite{ana13} \\
PSR J0437+4715   & 0.224   & 0.007   & 0.007   & \cite{rea16} \\
PSR J0751+1807   & 0.16    & 0.01    & 0.01    & \cite{dea16} \\
PSR J1141-6545   & 1.02    & 0.01    & 0.01    & \cite{bha08} \\
PSR J1145-6545   & 1.00    & 0.02    & 0.02    & \cite{vdH07} \\
PSR J1614-2230   & 0.493   & 0.003   & 0.003   & \cite{foa16} \\
PSR J1713+0747   & 0.286   & 0.012   & 0.012   & \cite{zha15} \\
PSR J1730+0333   & 0.181   & 0.007   & 0.005   & \cite{ana12} \\
PSR J1802-2124   & 0.78    & 0.04    & 0.04    & \cite{fea10} \\
PSR J1909-3744   & 0.214   & 0.003   & 0.003   & \cite{foa16} \\
PSR J1918-0642   & 0.219   & 0.012   & 0.011   & \cite{foa16} \\
PSR J2234+0611 B & 0.298   & 0.015   & 0.012   & \cite{sta18} \\ 
PTFEB11.441      & 0.54    & 0.05    & 0.05    & \cite{laa12} \\
QS Vir           & 0.7816  & 0.0130  & 0.0130  & \cite{paa16} \\
RR Cae           & 0.4475  & 0.0015  & 0.0015  & \cite{paa17} \\ 
\noalign{\smallskip}               
\hline                             
\end{tabular}                      
\end{center}                       
\end{table*}                       
\begin{table*}
\caption[par]{High-precision ($\Delta M/M<0.1$) masses and uncertainties for
$N=51$ white dwarf (WD) stars hosted in binary systems, and $N=25$ nearby
stars belonging to the 25-pc WD sample \cite{sua17}.  Masses are in solar
units.   Continuation of Table \ref{t:mawda}.}
\label{t:mawdb}
\begin{center}
\begin{tabular}{lllll}
\hline
\multicolumn{1}{c}{system} &
\multicolumn{1}{c}{$M\quad$} &
\multicolumn{1}{c}{$+\Delta M\quad$} &
\multicolumn{1}{c}{$-\Delta M\quad$} &
\multicolumn{1}{c}{ref} \\
\hline\noalign{\smallskip}
RX J0648.0-4418  & 1.28    & 0.05    & 0.05    & \cite{mea09} \\
SDSS J0024+1745  & 0.5340  & 0.0090  & 0.0090  & \cite{paa17} \\
SDSS J0106-0014  & 0.4406  & 0.0144  & 0.0144  & \cite{paa17} \\
SDSS J0110+1326  & 0.4656  & 0.0091  & 0.0091  & \cite{paa17} \\
SDSS J0138-0016  & 0.5290  & 0.0100  & 0.0100  & \cite{pab12} \\
SDSS J0314+0206  & 0.5964  & 0.0088  & 0.0088  & \cite{paa17} \\
SDSS J0857+0342  & 0.5140  & 0.0490  & 0.0490  & \cite{pac12} \\
SDSS J1021+1744  & 0.5338  & 0.0038  & 0.0038  & \cite{paa17} \\
SDSS J1028+0931  & 0.4146  & 0.0036  & 0.0036  & \cite{paa17} \\
SDSS J1123-1155  & 0.6050  & 0.0079  & 0.0079  & \cite{paa17} \\
SDSS J1210+3347  & 0.4150  & 0.0100  & 0.0100  & \cite{pya12} \\
SDSS J1212-0123  & 0.4393  & 0.0022  & 0.0022  & \cite{paa12} \\
SDSS J1307+2156  & 0.6098  & 0.0031  & 0.0031  & \cite{paa17} \\
SDSS J1329+1230  & 0.3916  & 0.0234  & 0.0234  & \cite{paa17} \\
SDSS J2235+1428  & 0.3977  & 0.0220  & 0.0220  & \cite{paa17} \\
Sirius B         & 1.018   & 0.011   & 0.011   & \cite{boa17} \\
Stein 2051 B     & 0.675   & 0.051   & 0.051   & \cite{sha17} \\
V471 Tau         & 0.8400  & 0.0500  & 0.0500  & \cite{OBS01} \\
WD 0000-345      & 0.68    &  0.03   & 0.03    & \cite{sua17} \\
WD 0034-602      & 0.95    &  0.04   & 0.04    & \cite{sua17} \\
WD 0046+051      & 0.67    &  0.02   & 0.02    & \cite{sua17} \\
WD 0136+152      & 0.66    &  0.02   & 0.02    & \cite{sua17} \\
WD 0243-026      & 0.73    &  0.03   & 0.03    & \cite{sua17} \\
WD 0310-688      & 0.65    &  0.04   & 0.04    & \cite{sua17} \\
WD 0311-649      & 0.29    &  0.02   & 0.02    & \cite{sua17} \\
WD 0423+044      & 0.72    &  0.04   & 0.04    & \cite{sua17} \\
WD 0457-004      & 1.10    &  0.01   & 0.01    & \cite{sua17} \\
WD 0511+079      & 0.84    &  0.03   & 0.03    & \cite{sua17} \\
WD 0548-001      & 0.70    &  0.02   & 0.02    & \cite{sua17} \\
WD 0552-041      & 0.90    &  0.02   & 0.02    & \cite{sua17} \\
WD 0644+025      & 0.99    &  0.01   & 0.01    & \cite{sua17} \\
WD 0659-063      & 0.63    &  0.03   & 0.03    & \cite{sua17} \\
WD 0738-172      & 0.61    &  0.03   & 0.03    & \cite{sua17} \\
WD 0816-310      & 0.71    &  0.06   & 0.06    & \cite{sua17} \\
\noalign{\smallskip}
\hline                             
\end{tabular}                      
\end{center}                       
\end{table*}                       
\begin{table*}
\caption[par]{High-precision ($\Delta M/M<0.1$) masses and uncertainties for
$N=51$ white dwarf (WD) stars hosted in binary systems, and $N=25$ nearby
stars belonging to the 25-pc WD sample \cite{sua17}.  Masses are in solar
units.   Continuation of Table \ref{t:mawdb}.}
\label{t:mawdc}
\begin{center}
\begin{tabular}{lllll}
\hline
\multicolumn{1}{c}{system} &
\multicolumn{1}{c}{$M\quad$} &
\multicolumn{1}{c}{$+\Delta M\quad$} &
\multicolumn{1}{c}{$-\Delta M\quad$} &
\multicolumn{1}{c}{ref} \\
\hline\noalign{\smallskip}
WD 0856-007      & 0.64    &  0.03   & 0.03    & \cite{sua17} \\
WD 1036-204      & 0.78    &  0.03   & 0.03    & \cite{sua17} \\
WD 1105-340      & 0.69    &  0.04   & 0.04    & \cite{sua17} \\
WD 1236-495      & 0.98    &  0.02   & 0.02    & \cite{sua17} \\
WD 1333+005      & 0.4356  & 0.0016  & 0.0016  & \cite{paa17} \\
WD 2047+372      & 0.82    &  0.02   & 0.02    & \cite{sua17} \\
WD 2159-754      & 0.97    &  0.04   & 0.04    & \cite{sua17} \\
WD 2211-392      & 0.80    &  0.04   & 0.04    & \cite{sua17} \\
WD 2341+322      & 0.62    &  0.01   & 0.01    & \cite{sua17} \\
WD 2359-434      & 0.83    &  0.02   & 0.02    & \cite{sua17} \\
40 Eri B         & 0.573   & 0.018   & 0.018   & \cite{MHM17} \\
\noalign{\smallskip}
\hline                             
\end{tabular}                      
\end{center}                       
\end{table*}                       
\begin{table*}
\caption[par]{High-precision ($\Delta M/M<0.1$) masses and uncertainties for
$N=37$
neutron star (NS) members in binary systems.  Masses are in solar units.}
\label{t:mans}
\begin{center}
\begin{tabular}{lllll}
\hline
\multicolumn{1}{c}{system} &
\multicolumn{1}{c}{$M\quad$} &
\multicolumn{1}{c}{$+\Delta M\quad$} &
\multicolumn{1}{c}{$-\Delta M\quad$} &
\multicolumn{1}{c}{ref} \\
\hline\noalign{\smallskip}
LMC X-4          & 1.57    & 0.11    & 0.11    & \cite{faa15} \\
PSR B1534+12 p   & 1.3330  & 0.0004  & 0.0004  & \cite{foa14} \\
PSR B1534+12 c   & 1.3455  & 0.0004  & 0.0004  & \cite{foa14} \\
PSR B1855+09     & 1.31    & 0.12    & 0.10    & \cite{foa16} \\
PSR B1913+16     & 1.4398  & 0.0002  & 0.0002  & \cite{wea10} \\
PSR B1913+16     & 1.3886  & 0.0002  & 0.0002  & \cite{wea10} \\
PSR B2127+11 p   & 1.358   & 0.010   & 0.010   & \cite{jaa06} \\
PSR B2127+11 c   & 1.354   & 0.010   & 0.010   & \cite{jaa06} \\
PSR B2303+46     & 1.34    & 0.10    & 0.10    & \cite{vKK99} \\
PSR J0337+1715   & 1.4378  & 0.0013  & 0.0013  & \cite{raa14} \\
PSR J0348+0432   & 2.01    & 0.04    & 0.04    & \cite{ana13} \\
PSR J0437+4715   & 1.44    & 0.07    & 0.07    & \cite{rea16} \\
PSR J0453+1559 p & 1.559   & 0.005   & 0.005   & \cite{maa15} \\
PSR J0453+1559 c & 1.174   & 0.004   & 0.004   & \cite{maa15} \\
PSR J0737-3039 A & 1.3381  & 0.007   & 0.007   & \cite{kra06} \\
PSR J0737-3039 B & 1.2489  & 0.007   & 0.007   & \cite{kra06} \\
PSR J0751+1807   & 1.64    & 0.15    & 0.15    & \cite{dea16} \\
PSR J1012+5307   & 1.83    & 0.11    & 0.11    & \cite{ana16} \\
PSR J1141-6545   & 1.27    & 0.01    & 0.01    & \cite{bha08} \\
PSR J1145-6545   & 1.28    & 0.02    & 0.02    & \cite{vdH07} \\
PSR J1614-2230   & 1.928   & 0.017   & 0.017   & \cite{foa16} \\
PSR J1713+0747   & 1.31    & 0.11    & 0.11    & \cite{zha15} \\
PSR J1738+0333   & 1.47    & 0.07    & 0.06    & \cite{ana12} \\
PSR J1756-2251 p & 1.341   & 0.007   & 0.007   & \cite{fea14} \\
PSR J1756-2251 c & 1.230   & 0.007   & 0.007   & \cite{fea14} \\
PSR J1757-1854 p & 1.3384  & 0.0009  & 0.0009  & \cite{caa18} \\
PSR J1757-1854 c & 1.3946  & 0.0009  & 0.0009  & \cite{caa18} \\
PSR J1802-2124   & 1.24    & 0.11    & 0.11    & \cite{fea10} \\
PSR J1807-2500B  & 1.3655  & 0.021   & 0.021   & \cite{lya12} \\
PSR J1829+2456 p & 1.30    & 0.05    & 0.05    & \cite{vdH07} \\
PSR J1829+2456 c & 1.27    & 0.11    & 0.07    & \cite{vdH07} \\
PSR J1903+0327   & 1.667   & 0.021   & 0.021   & \cite{fra11} \\
PSR J1906+0746   & 1.291   & 0.011   & 0.011   & \cite{vLa15} \\
PSR J1909-3744   & 1.55    & 0.03    & 0.03    & \cite{foa16} \\
PSR J1918-0642   & 1.19    & 0.10    & 0.09    & \cite{foa16} \\
PSR J2234+0611 A & 1.353   & 0.014   & 0.017   & \cite{sta18} \\
SMC X-1          & 1.21    & 0.12    & 0.12    & \cite{faa15} \\
\noalign{\smallskip}               
\hline                             
\end{tabular}                      
\end{center}                       
\end{table*}                       
\begin{table*}
\caption[par]{Masses and uncertainties for $N=17$ black hole (BH) stellar
remnants
hosted in binary systems.  The post-merger body from GW 170817 is also
included, where $-\Delta M=0.05=0.01$ (intrinsic error \cite{aba17}) + 0.04
(upper limit of ejected matter \cite{aia18}).  Masses are in solar units.}
\label{t:mabh}
\begin{center}
\begin{tabular}{lllll}
\hline
\multicolumn{1}{c}{system} &
\multicolumn{1}{c}{$M\quad$} &
\multicolumn{1}{c}{$+\Delta M\quad$} &
\multicolumn{1}{c}{$-\Delta M\quad$} &
\multicolumn{1}{c}{ref} \\
\hline\noalign{\smallskip}
A0620-00         &  6.61   & 0.23    & 0.17    & \cite{GRC14} \\
Cyg X-1          & 10.05   & 3.20    & 3.20    & \cite{oro03} \\
GRO J0422+32     &  3.97   & 0.95    & 0.95    & \cite{GeH03} \\
GRO J1655-40     &  6.30   & 0.27    & 0.27    & \cite{oro03} \\
GRS 1915+105     & 14      & 4       & 4       & \cite{GCM01} \\
GS 1124-683      & 11.0    & 2.1     & 1.4     & \cite{wua16} \\
GS 2000+25       &  7.46   & 0.32    & 0.32    & \cite{oro03} \\
GS 2023+338      & 11.72   & 1.66    & 1.66    & \cite{oro03} \\
H 1705-250       &  6.97   & 1.33    & 1.33    & \cite{oro03} \\
LMC X-1          & 10.91   & 1.54    & 1.54    & \cite{ora09} \\
LMC X-3          &  7.55   & 1.62    & 1.62    & \cite{oro03} \\
M33 X-7          & 15.65   & 1.45    & 1.45    & \cite{ora07} \\
NGC 3201         &  4.36   & 0.41    & 0.41    & \cite{gib18} \\
SAX J1819.3-2525 &  7.12   & 0.30    & 0.30    & \cite{oro03} \\
XTE J1118+480    &  7.46   & 0.34    & 0.69    & \cite{GRC14} \\
XTE J1550-564    &  9.10   & 0.61    & 0.61    & \cite{ora11} \\
4U 1543-47       &  9.42   & 0.97    & 0.97    & \cite{oro03} \\
GW 170817        &  2.74   & 0.04    & 0.05    & \cite{aba17} \\
\noalign{\smallskip}
\hline                             
\end{tabular}                      
\end{center}                       
\end{table*}                       

\end{document}